\documentclass[journal]{IEEEtran}
\usepackage{epsf, psfrag, amssymb, amsfonts, amsmath, cite,enumerate}
\usepackage{graphicx, subfigure, color,bbm}
\usepackage{boxedminipage}
\usepackage{multirow}
\usepackage{booktabs}
\usepackage{algorithm, algorithmic}
\usepackage{dblfloatfix}

\newtheorem{remark}{Remark}

\newtheorem{theorem}{Theorem}
\newtheorem{example}{Example}
\newtheorem{corollary}{Corollary}
\newtheorem{lemma}{Lemma}

\def\psfancypar#1#2{\begingroup\def\par{\endgraf\endgroup\lineskiplimit=0pt}
               \setbox2=\hbox{\large\sc #2}
               \newdimen\tmpht \tmpht \ht2 \advance\tmpht by \baselineskip
               \font\hhuge=Times-Bold at \tmpht
               \setbox1=\hbox{{\hhuge #1}}
               \count7=\tmpht \count8=\ht1
               \divide\count8 by 1000 \divide\count7 by \count8
               \tmpht=.001\tmpht\multiply\tmpht by \count7
               \font\hhuge=Times-Bold at \tmpht
               \setbox1=\hbox{{\hhuge #1}}
               \noindent
                \hangindent1.05\wd1
               \hangafter=-2 {\hskip-\hangindent
               \lower1\ht1\hbox{\raise1.0\ht2\copy1}%
                \kern-0\wd1}\copy2\lineskiplimit=-1000pt}

\newcommand{\beq}{\begin{equation}}
\newcommand{\eeq}{\end{equation}}
\newcommand{\bqa}{\begin{eqnarray}}
\newcommand{\eqa}{\end{eqnarray}}
\newcommand{\bqn}{\begin{eqnarray*}}
\newcommand{\eqn}{\end{eqnarray*}}
\newcommand{\nn}{\nonumber}

\newcommand{\be}{\begin{enumerate}}
\newcommand{\ee}{\end{enumerate}}
\newcommand{\bi}{\begin{itemize}}
\newcommand{\ei}{\end{itemize}}
\newcommand{\bd}{\begin{description}}
\newcommand{\ed}{\end{description}}
\newcommand{\ba}{\begin{array}}
\newcommand{\ea}{\end{array}}
\newcommand{\bde}{\begin{definition}}
\newcommand{\ede}{\end{definition}}
\newcommand{\bex}{\begin{example}}
\newcommand{\eex}{\end{example}}

\def\boxit#1{\vbox{\hrule\hbox{\vrule\kern3pt
        \vbox{\kern3pt#1\kern3pt}\kern3pt\vrule}\hrule}}

\def\reals{ { {\rm  I \kern-0.15em R }  } }
\def\complex{ {\,{{\rm C} \kern-0.50em \raise0.20ex {  |}}\, }}

\def\0bf{{\bf 0}}
\def\1bf{{\bf 1}}
\def\2bf{{\bf 2}}
\def\3bf{{\bf 3}}
\def\4bf{{\bf 4}}
\def\5bf{{\bf 5}}
\def\6bf{{\bf 6}}
\def\7bf{{\bf 7}}
\def\8bf{{\bf 8}}
\def\9bf{{\bf 9}}

\def\Rbf{{\bf R}}

\def\Rxx{\Rbf_{\ssstyle X\kern-.1em X}}

\let\ssstyle=\scriptscriptstyle

\def\Kout{\setbox1=\hbox{\Huge\bf K}\hbox to
1.05\wd1{\hspace{.05\wd1}
}}
\def\Sout{\setbox1=\hbox{\Huge\bf S}\hbox to 1.05\wd1{\hspace{.05\wd1}
}}

\usepackage{bm}
\newcommand*{\B}[1]{\ifmmode\bm{#1}\else\textbf{#1}\fi}
\DeclareMathOperator*{\argmin}{arg\,min}

\begin{document}
%
\title{Quantized Consensus by the ADMM: Probabilistic versus Deterministic Quantizers}
\author{Shengyu Zhu,~\IEEEmembership{Student~Member,~IEEE,} and Biao Chen,~\IEEEmembership{Fellow,~IEEE}\thanks{This work was supported by National Science Foundation under Award CCF1218289,  by Army Research Office under Award W911NF-12-1-0383, and Air Force Office of Scientific Research under Award FA9550-10-1-0458. The material of this paper was presented in part at IEEE GlobalSIP 2015 \cite{Zhu2015a}.}\thanks{S. Zhu and B. Chen are with the Department of Electrical Engineering and Computer Science, Syracuse University, Syracuse, NY 13244 USA (e-mail: szhu05@syr.edu; bichen@syr.edu).}}

\maketitle

\begin{abstract}
This paper develops efficient algorithms for distributed average consensus with quantized communication using the alternating direction method of multipliers (ADMM). We first study the effects of probabilistic and deterministic quantizations on a distributed ADMM algorithm. With probabilistic quantization, this algorithm yields linear convergence to the desired average in the mean sense with a bounded variance. When deterministic quantization is employed, the distributed ADMM either converges to a consensus or cycles with a finite period after a finite-time iteration. In the cyclic case, local quantized variables have the same mean over one period and hence each node can also reach a consensus. We then obtain an upper bound on the consensus error which depends only on the quantization resolution and the average degree of the network. Finally, we propose a two-stage algorithm which combines both probabilistic and deterministic quantizations. Simulations show that the two-stage algorithm, without picking small algorithm parameter, has consensus errors that are typically less than {one} quantization resolution for {all} connected networks where agents' data can be of {arbitrary magnitudes}. 
\end{abstract}
\begin{IEEEkeywords}
Quantized consensus, dither, probabilistic quantization, deterministic quantization, alternating direction method of multipliers, linear convergence.
\end{IEEEkeywords}

\section{Introduction}
\IEEEPARstart{I}{n} recent years there has been considerable interest in distributed average consensus where a group of agents aim to reach a consensus on the average of their measurements \cite{Ren2007,Cao2013,Lynch1996distributed, Ren2005,Xiao2005,Xu1996,Kashyap2007,Xiao2004, Jakovetic2010,Nedic2009,Aysal2009,Boyd2006,Schizas2008,Zhu2009,Erseghe2011,Aysal2008,Kar2010,Chamie2014,Carli2010}. This is largely motivated by numerous applications in control, signal processing, and computer science. For example, the distributed averaging is a fundamental problem in {\it{ad hoc}} network applications, such as distributed agreement and synchronization \cite{Lynch1996distributed}, distributed coordination of mobile autonomous agents \cite{Ren2005}, and distributed data fusion in sensor networks \cite{Xiao2005}. It has also found applications in load balancing for parallel computers \cite{Xu1996}.

We consider in this paper distributed averaging algorithms where nodes only exchange information with their immediate neighbors. These algorithms are extremely attractive for large scale networks characterized by the lack of centralized access to information. They are also energy efficient and enhance the survivability of the networks, compared with fusion center based processing. However, a number of factors such as limited bandwidth, sensor battery power, and computing resources place tight constraints on the rate and form of information exchange amongst neighboring nodes, resulting in {\em quantized consensus} problems \cite{Ren2007,Kashyap2007}. This paper is specifically devoted to developing efficient algorithms for quantized consensus in connected networks with static topologies.
\subsection{Related work}
There are three widely used methods for solving distributed averaging problems. A classical approach is to update the state of each node with a weighted average of values from neighboring nodes \cite{Xiao2004,Jakovetic2010,Nedic2009}. The matrix, consisting of the weights associated with the edges, is chosen to be doubly stochastic to ensure convergence to the average. Another method is a gossip based algorithm, initially introduced in \cite {Tsitsiklis1984} for consensus problems and further studied in \cite{Kashyap2007,Aysal2009, Boyd2006}, among others. The third approach is to employ the ADMM which is an iterative algorithm for solving convex problems and has received much attention recently (see \cite{BoydADMM} and references therein). The idea is to formulate the data average as the solution to a least-squares problem and manipulate the ADMM updates to derive a distributed algorithm  \cite{Schizas2008, Zhu2009, Erseghe2011}.

In the most ideal case where agents are able to send and receive real values with infinite precision, the three methods all lead to the desired consensus at the average. When quantization is imposed, however, these methods do not directly apply. A well studied approach for quantized consensus is to use dithered quantizers which add noises to agents' variables before quantization\cite{Schuchman1964}. By imposing certain conditions, the quantization error sequence becomes independent and identically distributed (i.i.d.) and is also independent of the input sequence. The classical approach and the gossip based algorithm then yield the almost sure consensus at a common but random quantization level with the expectation of the consensus value equal to the desired average  \cite{Aysal2008,Kar2010,Carli2010}. To the best of our knowledge, there have been no existing results on the ADMM based method for quantized consensus. Nevertheless, since the quantization error of dithered quantizer is zero-mean and has a bounded variance, we can immediately extend the results in \cite{Zhu2009, Erseghe2011} to quantized consensus (see Section \ref{sec:PQ}). That is, the ADMM based method using dithered quantization leads to the consensus at the data average in the mean sense whose variance converges to a finite value. 

Meanwhile, studies on distributed average consensus with deterministic quantizers have been scarcely reported. Deterministic quantization makes the problem much harder to deal with as the error terms caused by quantization no longer possess tractable statistical characteristics \cite{Aysal2008,Kar2010}. The authors in \cite{Nedic2009} show that the classical approach, where a quantization rule that rounds the values down is adopted, converges to a consensus with an error from the average depending on the quantization resolution, the number of agents, the agents' data and the updated weights of each agent.  A recent result of \cite{Chamie2014} indicates that this approach, with appropriate choices of the weights, reaches a quantized consensus close to the average in finite time or leads all agents' variables to cycle in a small neighborhood around the average; in the latter case, however, the consensus is not guranteed. The gossip based algorithms in \cite{Carli2010} and \cite{Kashyap2007} have similar results to those of the classical approach. The ADMM based algorithms for deterministically quantized consensus, however, have not yet been explored.

\subsection{Our contributions}
One shall note that the consensus error for deterministically quantized consensus in \cite{Nedic2009,Carli2010} is much undesired when the number of agents or the range of agents' data becomes very large. Unfortunately, this is typically the case in large scale networks or big data settings. The ADMM has been known to be an efficient algorithm for large scale optimizations and used in various applications such as regression and classification \cite{BoydADMM}. Moreover, \cite{He2012, Hong2012, Deng2016} validate the fast convergence of the ADMM and \cite{Zhu2009, Erseghe2011} demonstrate the resilience of the ADMM to noise, link failures, etc. We therefore expect ADMM based methods to work well for quantized consensus problems, with regards to both the consensus error and the convergence time.

We first study the effect of probabilistic quantization \cite{Xiao2005a}, which is equivalent to a dithering method as shown by \cite[Lemma 2]{Aysal2008}, on the ADMM based method. Utilizing the first and second order moments of the probabilistic quantizer output, we establish the convergence to the average in the mean sense based on existing convergence results of the ADMM. Furthermore, recent work of \cite{Shi2014} enables us to immediately characterize the convergence rate of the distributed ADMM with probabilistic quantization.

The main contribution of this paper is to design and analyze an ADMM based approach using deterministic quantization. We establish that a distributed deterministically quantized ADMM algorithm either converges to a consensus or cycles around the average after a finite-time iteration as long as a mild initialization condition is satisfied. We also show that the cyclic period is finite and that the quantized variable at each node has the same mean over one period. Thus, a consensus can be reached within finite iterations for both convergent and cyclic cases. We then derive an upper bound that only depends on the quantization resolution and the average degree of the undirected graph (two times the ratio of the number of edges to the number of nodes). This is much preferred for large scale networks as it does not rely on the number of agents or the agent's data.

While numerical examples show that the deterministically quantized ADMM converges in most cases, we notice that it may reach different consensus values with different initial variable values. It is well known that a good starting point usually helps in such settings. This inspires our approach for quantized consensus which first uses the probabilistic method to obtain a good starting point and then employs the deterministic algorithm. Simulations show that this two-stage approach tends to converge and also performs best among all existing methods using deterministic quantization in terms of the consensus error. 

\subsection{Paper organization}
The rest of this paper is organized as follows. Section \ref{sec:ADMMnoQ} reviews the application of the ADMM to the distributed averaging problem without quantization, which leads to a distributed ADMM algorithm. We then develop several convergence results of this algorithm; they will be used later to establish our main results. Section \ref{sec:problemformulation} defines probabilistic and deterministic quantization schemes. Their effects on the distributed ADMM are studied respectively in Sections \ref{sec:PQ} and \ref{sec:DQ}. Section \ref{sec:algorithm} describes the proposed algorithm for quantized consensus which combines the two quantized ADMM methods, followed by simulation results in Section \ref{sec:simulation}. Section \ref{sec:conclusion} concludes the paper. 
\subsection{Notations}
Denote by $\|\bm{x}\|_2$ the Euclidean norm of a vector $\bm x$ and $\langle \bm{x},\bm{y} \rangle$ the inner product of two vectors $\bm x$ and $\bm y$. Given a semidefinite matrix $\bm G$ with proper dimensions, the $\bm G$-norm of $\bm x$ is $\|\bm x\|_{\bm G}=\sqrt{\bm{x}^T\bm{Gx}}$. Also denote $\sigma_{\max}(\bm{M})$ as the largest singular value of a square matrix $\bm M$ and $\tilde{\sigma}_{\min}(\bm{M})$ as the smallest nonzero singular value of $\bm{M}$. 

We use two definitions of rate of convergence for an iterative algorithm. A sequence $\bm x^k$, where the superscript $k$ stands for time index, is said to converge \emph{Q-linearly} to a point $\bm x^*$ if there exists a number $\upsilon\in(0,1)$ such that $\lim_{k\to\infty}\frac{\|\bm x^{k+1}-\bm x^*\|}{\|\bm x^k-\bm x^*\|}=\upsilon$ with $\|\cdot\|$ being a vector norm. A sequence $\bm y^k$ is said to converge \emph{R-linearly} to $\bm y^*$ if for all $k$, $\|\bm y^k-\bm y^*\|\leq\|\bm x^k-\bm x^*\|$ where $\bm x^k$ converges Q-linearly to $\bm x^*$.

\section{Distributed Average Consensus by the ADMM}
\label{sec:ADMMnoQ}
This section introduces the consensus ADMM (CADMM) for average consensus without quantization. This ideal case provides a good understanding of how the ADMM works for distributed average consensus. We start with the setting of the distributed averaging problem.

\subsection{Problem setting}
\label{sec:orgnotation}
Consider a connected network of $N$ agents which are bidirectionally connected by $E$ edges (and thus $2E$ arcs). We describe this network as a symmetric directed graph $\mathcal{G}_d=\{\mathcal{V},\mathcal{A}\}$ or an undirected graph $\mathcal{G}_u=\{\mathcal{V},\mathcal{E}\}$, where $\mathcal{V}$ is the set of vertices with cardinality $|\mathcal{V}|=N$, $\mathcal{A}$ is the set of arcs with $|\mathcal{A}|=2E$ and $\mathcal{E}$ is the set of edges with $|\mathcal{E}|=E$. Assume that the topology of the network is fixed throughout this paper. Let $r_i\in\mathbb{R}$ be the local data only available at node $i$, $i=1,2,\cdots,N$, and $\bm{r}\in\mathbb{R}^N$ the vector concatenating all $r_i$. The goal of distributed average consensus is to compute the data average  
\begin{equation}
\label{eqn:averageconsensus}
\begin{aligned}
x_\text{avg}=\frac{1}{N}\sum_{i=1}^N r_i
\end{aligned}
\end{equation} 
by local data exchanges among neighboring nodes.

\subsection{Application of the ADMM to distributed average consensus: CADMM}
The ADMM applies in general to the convex optimization problem in the form of 
\begin{equation}
\begin{aligned}
\label{eqn:admm}
& \underset{\bm y_1,\bm y_2}{\text{minimize}}
& & g_1(\bm y_1)+g_2(\bm y_2)\\
& \text{subject to}
& & \bm C_1\bm y_1+\bm C_2\bm y_2 = \bm c,
\end{aligned}
\end{equation}
where $\bm y_1$ and $\bm y_2$ are optimization variables, $g_1$ and $g_2$ are convex functions, and $\bm C_1\bm y_1+\bm C_2\bm y_2=\bm c$ is a linear constraint on $\bm y_1$ and $\bm y_2$. The ADMM solves a sequence of subproblems involving $g_1$ and $g_2$ one at a time and iterate to converge when, e.g., $g_1$ and $g_2$ are proper closed convex functions and the Lagrangian of (\ref{eqn:admm}) has a saddle point \cite{BoydADMM}. 

To apply the ADMM, we first formulate (\ref{eqn:averageconsensus}) as a convex optimization problem
\begin{equation}
\label{eqn:squareformulation}
\begin{aligned}
x_\text{avg}=\argmin_{\tilde{x}}\sum_{i=1}^N\frac{1}{2}(\tilde{x}-r_i)^2,
\end{aligned}
\end{equation} 
that is, the data average is the solution to a least-squares minimization problem. We continue to reformulate (\ref{eqn:squareformulation}) in the form of (\ref{eqn:admm}) as
\begin{equation}
\begin{aligned}
\label{eqn:admmformulation}
& \underset{\{x_i\},\{z_{ij}\}}{\text{minimize}}
& & \sum_{i=1}^N \frac{1}{2}(x_i-r_i)^2\\
& \text{subject to}
& & x_i=z_{ij},x_j=z_{ij},\forall (i,j)\in\mathcal{A},
\end{aligned}
\end{equation}
where $x_i$ is the local copy of the common optimization variable $\tilde{x}$ at node $i$ and $z_{ij}$ is an auxiliary variable imposing the consensus constraint on neighboring nodes $i$ and $j$. We emphasize that throughout the entire paper, $r_i$ represents the local data, i.e., the observation at the $i$th agent, while $x_i$ is referred to as the local variable. Since the network is connected, this constraint ensures the consensus to be achieved over the entire network, i.e., $x_i=x_j, \forall (i,j) \in \mathcal{A}$, which in turn guarantees the solution to (\ref{eqn:admmformulation}) is the data average $x_\text{avg}$. Further define $\bm x\in\mathbb{R}^{N}$ as a vector concatenating all $x_i$, $\bm z\in\mathbb{R}^{2E}$ as a vector concatenating all $z_{ij}$, and 
\begin{align}
\label{eqn:objectivefunction}
f(\bm x)=\frac{1}{2}\|\bm x-\bm r\|_2^2.
\end{align}
Then (\ref{eqn:admmformulation}) can be written in a matrix form as 
\begin{equation}
\label{eqn:matrixform}
\begin{aligned}
& \underset{x,z}{\text{minimize}}
& & f(\bm x)+g(\bm z)\\
& \text{subject to}
& & \bm{Ax}+\bm{Bz}=\bm 0,
\end{aligned}
\end{equation}
where $g(\bm z)=0$, and $\bm 0$ is a column vector with proper dimensions and all entries being $0$. Here $\bm B = [-\bm I_{2E}; -\bm I_{2E}]$ with $\bm I_{2E}$ being a $2E\times 2E$ identity matrix and $\bm A = [\bm A_1;\bm A_2]$ with $\bm A_1,\bm A_2\in\mathbb{R}^{2E\times N}$. If $(i,j)\in\mathcal{\bm A}$ and $z_{ij}$ is the $q$th entry of $\bm z$, then the $(q,i)$th entry of $\bm A_1$ and the $(q,j)$th entry of $\bm A_2$ are $1$; otherwise the corresponding entries are $0$. 

We are now ready to apply the ADMM to solve the consensus problem. The augmented Lagrangian of (\ref{eqn:matrixform}) is 
\begin{equation}
\label{eqn:Lagragian}
\begin{aligned}
L_\rho(\bm x,\bm z,\bm \lambda)=f(\bm x)+\langle \bm \lambda,\bm{Ax}+\bm{Bz}\rangle+\frac{\rho}{2}\|\bm{Ax}+\bm{Bz}\|_2^2,
\end{aligned}
\end{equation}
where $\bm \lambda=[\bm \beta; \bm \gamma]$ with $\bm \beta,\bm \gamma\in\mathbb{R}^{2E}$ is the Lagrange multiplier and $\rho$ is a positive algorithm parameter. At iteration $k+1$, the ADMM first obtains $\bm x^{k+1}$ by minimizing $L_\rho(\bm x,\bm z^k,\bm \lambda^k)$, then calculates $\bm z^{k+1}$ by minimizing $L_\rho(\bm x^{k+1},\bm z,\bm \lambda^k)$ and finally updates $\bm \lambda^{k+1}$ using $\bm x^{k+1}$ and $\bm z^{k+1}$. The updates are 
\begin{equation}
\label{eqn:admmupdates}
\begin{aligned}
\bm x\text{-update}:&~~~~~\nabla f(\bm x^{k+1}) + \bm A^T\bm\lambda^k\\&~+\rho \bm A^T(\bm A\bm x^{k+1}+\bm B\bm z^k)=\bm 0,\\
\bm z\text{-update}:&~~\bm B^T\bm \lambda^k+\rho \bm B^T(\bm A\bm x^{k+1}+\bm B\bm z^{k+1}) = \bm 0,\\
\bm \lambda\text{-update}:&~~\bm \lambda^{k+1}-\bm\lambda^k-\rho(\bm A\bm x^{k+1}+\bm B\bm z^{k+1})=\bm 0,
\end{aligned}
\end{equation} 
where $\nabla f(\bm x^{k+1})=\bm x^{k+1}-\bm r$ is the gradient of $f$ at $\bm x^{k+1}$. 

A nice property of the ADMM, known as \emph{global convergence}, states that the sequence $(\bm x^k,\bm z^k,\bm \lambda^k)$ generated by (\ref{eqn:admmupdates}) has a single limit point  $(\bm x^*,\bm z^*,\bm \lambda^*)$ which is a primal-dual solution to (\ref{eqn:Lagragian}). Proofs can be found in \cite{BoydADMM,Deng2016,He2012}. Noting that our objective function $f(\bm x)$ given in (\ref{eqn:objectivefunction}) is strongly convex in $\bm x$, we obtain $\bm x^*=\bm 1x_\text{avg}$ as the unique primal solution where $\bm 1$ denotes the $N$-dimensional column vector with all entries being $1$. To summarize, we have
\begin{lemma}[Global convergence of the ADMM \cite{BoydADMM,Deng2016,He2012}]
\label{lem:globalconvergence}
For any initial values $\bm x^0\in\mathbb{R}^N$, $\bm z^0\in\mathbb{R}^{2E}$ and $\bm \lambda^0\in\mathbb{R}^{4E}$, the updates in (\ref{eqn:admmupdates}) yield that as $k\to\infty$, $$\bm x^k\to \bm x^*, ~\bm z^k\to\bm z^*,~\text{and}~\bm \lambda^k\to\bm \lambda^*,$$
where $(\bm x^*,\bm z^*,\bm \lambda^*)$ is a primal-dual solution to (\ref{eqn:Lagragian}) and $\bm x^*=\bm 1x_\text{avg}$ is unique for the distributed average consensus problem (\ref{eqn:squareformulation}).
\end{lemma}

While (\ref{eqn:admmupdates}) provides an efficient centralized algorithm to solve (\ref{eqn:squareformulation}), it is not clear whether (\ref{eqn:admmupdates}) can be carried out in a distributed manner, i.e., data exchanges only occur within neighboring nodes. Interestingly, Lemma \ref{lem:globalconvergence} states that convergence for the ADMM is guaranteed regardless of initial values $\bm x^0,\bm z^0$ and $\bm \lambda^0$; there indeed exist initial values that decentralize (\ref{eqn:admmupdates}). Define $\bm M_+=\bm A_1^T+\bm A_2^T$ and $\bm M_-=\bm A_1^T-\bm A_2^T$ which are respectively the unoriented and oriented incidence matrices with respect to the directed graph $\mathcal{G}_d$. Initialize $\bm \beta^0= -\bm\gamma^0$ and $\bm z^0=\frac{1}{2}\bm M_+^T\bm x^0$. As shown in \cite{Shi2014}, the updates in (\ref{eqn:admmupdates}) lead to 
\begin{equation}
\label{eqn:distributedversiond_plug}
\begin{aligned}
&x_i^{k+1}=\frac{1}{1+2\rho|\mathcal{N}_i|}\Bigg(\rho|\mathcal{N}_i|x_i^k+\rho\sum_{j\in\mathcal{N}_i} x_j^k-\alpha_i^k+r_i\Bigg),\\
&\alpha_i^{k+1}=\alpha_i^k+\rho\Bigg(|\mathcal{N}_i|x_i^{k+1}-\sum_{j\in\mathcal{N}_i} x_j^{k+1}\Bigg)
\end{aligned}
\end{equation} 
at node $i$, where $\mathcal{N}_i$ denotes the set of neighbors of node $i$ and $\alpha_i^k$ is the $i$th entry of $\bm \alpha^k=\bm M_-\bm \beta^k\in\mathbb{R}^N$. Obviously, (\ref{eqn:distributedversiond_plug}) is fully decentralized as the updates of $x_i^{k+1}$ and $\alpha_i^{k+1}$ only rely on local and neighboring information. Therefore (\ref{eqn:distributedversiond_plug}) can be used for distributed average consensus. We refer to (\ref{eqn:distributedversiond_plug}) as the CADMM for distributed average consensus.

If we further initialize $\bm\beta^0$ in the column space of $\bm M_-^T$ (e.g., $\bm \beta^0=\bm 0$), then $\bm \beta^k$ lies in the column space of $\bm M_-^T$ and converges to a unique $\bm\beta^*$. We will use this result immediately but postpone its proof to Lemma \ref{lem:linearconvergence}. Note that this implies $\bm\alpha^k$ in (\ref{eqn:distributedversiond_plug}) converges uniquely to $\bm\alpha^*=\bm M_-\bm\beta^*$. We also notice an interesting relation between $\bm\alpha$ and $\bm\beta$ even though $\bm M_-\bm M_-^T$ is rank deficient.\footnote{As defined in Section \ref{sec:LCofDCADMM}, $\bm M_-\bm M_-^T=2\bm L_-$ where $\bm L_-$ is the signed Laplacian matrix of the connected undirected graph and always has $0$ as its eigenvalue. See \cite{ChungSpectral}.}
 
\begin{lemma}
\label{lem:abrelation}
Given a connected network, if $\bm\beta$ lies in the column space of $\bm M_-^T$, then $\bm \alpha$ and $\bm \beta$ are one-to-one correspondence, i.e., for $\bm\alpha = \bm M_-\bm\beta$ and $\bm\alpha'=\bm M_-\bm\beta'$ where $\bm\beta$ and $\bm\beta'$ are in the column space of $\bm M_-^T$, $\bm\alpha = \bm\alpha'$ if and only if $\bm\beta=\bm\beta'$.
\end{lemma}
\begin{IEEEproof}
That $\bm\beta=\bm\beta'$ implying $\bm\alpha=\bm\alpha'$ is straightforward. Consider $\bm\alpha=\bm M_-\bm\beta$ and write $\bm\beta=\bm M_-^T\bm b$  for some $\bm b\in\mathbb{R}^N$. $\bm\alpha', \bm\beta'$ and $\bm b'$ are similarly defined. Then we have 
\begin{align}
\|\bm\alpha-\bm\alpha'\|_2 
&=\|\bm M_-\bm M_-^T(\bm b-\bm b')\|_2\nn\\
&\geq\tilde{\sigma}_{\min}(\bm M_-)\|\bm M_-^T(\bm b-\bm b')\|_2\nn\\
&=\tilde{\sigma}_{\min}(\bm M_-)\|\bm\beta-\bm\beta'\|_2,\nn
\end{align}
where $\tilde{\sigma}_{\min}(\bm M_-)$ is the smallest nonzero singular value of $\bm M_-$, whose existence is guaranteed for a connected graph \cite{ChungSpectral}. We therefore have $\bm\beta=\bm\beta'$ if $\bm\alpha=\bm\alpha'$.
\end{IEEEproof}

It is therefore meaningful to define an \emph{initialization condition} for the CADMM. A similar global convergence property for the CADMM is given in Lemma \ref{lem:linearconvergence_DCADMM}.
\begin{center}
{\fbox{\begin{minipage}{0.9\linewidth}
{\bf Initialization condition for the CADMM}: $\bm x^0$ can be any vector in $\mathbb{R}^N$ and $\bm\alpha^0$ lies in the column space of $\bm M_-\bm M_-^T$.
\end{minipage}}}
\end{center}

\begin{lemma}[Global convergence of the CADMM]
\label{lem:linearconvergence_DCADMM}
For any $\bm x^0$ and $\bm \alpha^0$ satisfying the initialization condition, the CADMM leads to $$\bm x^k\to \bm x^*~\text{and}~\bm\alpha^k\to\bm\alpha^*~\text{as}~k\to\infty,$$
where $\bm x^*=\bm 1x_\text{avg}$ and $\bm \alpha^*=\bm r -\bm 1 x_{\text{avg}}$ which lies in the column space of $\bm M_-\bm M_-^T$ are both unique.
\end{lemma}
\begin{IEEEproof}
Global convergence follows from Lemmas \ref{lem:globalconvergence} and \ref{lem:abrelation} together with the fact that $\bm\beta^k$ converges to a unique $\bm\beta^*$ which lies in the column space of $\bm M_-^T$.

Now taking $k\to\infty$ in (\ref{eqn:distributedversiond_plug}) and using the fact that $x_i^*=x_\text{avg}$ for $i=1,2,\cdots,N$, we have $$\alpha_i^*=r_i-x_\text{avg}.$$
\end{IEEEproof}

Throughout the rest of this paper, we assume that the CADMM, wherever used, is initialized to satisfy the initialization condition.

\subsection{Linear convergence of the CADMM}
\label{sec:LCofDCADMM}
We investigate two properties of the CADMM; the first property is built on global convergence while the second considers the rate of convergence. 

Define $\bm L_+=\frac{1}{2}\bm M_+\bm M_+^T$ and $\bm L_-=\frac{1}{2}\bm M_-\bm M_-^T$ which are respectively the signless and signed Laplacian matrices with respect to $\mathcal{G}_u$. Let $\bm W\in \mathbb{R}^{N\times N}$ be the degree matrix related to the underlying network, i.e., a diagonal matrix with its $(i,i)$th entry being the degree of node $i$ and other entries being $0$. Then $\bm W = \frac{1}{2}(\bm L_++\bm L_-)$  and Lemma \ref{lem:abrelation} is an immediate result from the property of $\bm L_-$ \cite{ChungSpectral}. We rewrite (\ref{eqn:distributedversiond_plug}) in the matrix form as
\begin{equation}
\label{eqn:idealdcadmm}
\begin{aligned}
&\bm x^{k+1}=(\bm I_N+2\rho \bm W)^{-1}( \rho \bm L_+ \bm x^k-\bm \alpha^k +\bm r),\\\nn
&\bm\alpha^{k+1} = \bm\alpha^k +\rho \bm L_-\bm x^{k+1},\nn
\end{aligned}
\end{equation}
or equivalently,
\begin{equation}
\label{eqn:iterateform}
\begin{aligned}
\bm s^{k+1}=\bm D\bm s^k,
\end{aligned}
\end{equation}
with 
$$\bm s^k = \left[\begin{array}{c} \bm x^{k}\\\bm\alpha^k\\\bm r{^k}\end{array}\right],$$and
\begin{equation}
\label{eqn:matrix_iterate}
\begin{aligned}
\bm D =\begin{bmatrix} \rho \bm D_0\bm L_+ & -\bm D_0& \bm D_0 \\ \rho^2 \bm L_-\bm D_0\bm L_+ & \bm I_N-\rho \bm L_-\bm D_0& \rho \bm L_-\bm D_0 \\\bm 0_N & \bm 0_N & \bm I_N \end{bmatrix},
\end{aligned}
\end{equation}
where $\bm 0_N$ denotes the $N\times N$ matrix with all entries being $0$, $\bm D_0=(\bm I_N+2\rho \bm W)^{-1}$, $\bm r^0=\bm r$, and hence, $\bm r^k=\bm r$. From (\ref{eqn:iterateform}), we have 
\begin{equation}
\begin{aligned}
\bm s^{k}=\bm D^k\bm s^0.\nn
\end{aligned}
\end{equation}
It is thus interesting to investigate how $\bm D^k$ behaves as $k\to\infty$. From (\ref{eqn:matrix_iterate}), a logical approach is to study $\bm D^k$ through the structures of $\bm L_-,\bm L_+$ and $\bm W$; fortunately, the global convergence property of the CADMM provides a simple argument to obtain a rough estimate of $\bm D^\infty$, which, nevertheless, is good enough for our purpose in establishing the main results. Note that we also have $\bm D^*=\bm D^\infty$ and $\bm s^*=[\bm x^*;\bm \alpha^*;\bm r^*]=[\bm x^\infty;\bm\alpha^\infty;\bm r^\infty]=\bm s^\infty$ as our optima due to global convergence of the CADMM. Our result about $\bm D^*$ is given below.
\begin{theorem}
\label{thm:Dproperty}
Consider $\bm D$ defined in (\ref{eqn:matrix_iterate}). Then
$$\bm D^* =\begin{bmatrix} \bm D_{11} & \bm D_{12} &\bm D_{13}\\\bm D_{21} &\bm D_{22} & \bm D_{23}\\\bm D_{31} & \bm D_{32} &\bm D_{33}\end{bmatrix}= \begin{bmatrix} \bm 0_N & \bm a_1\bm 1^T&\frac{1}{N}\bm 1\bm 1^T \\ \bm 0_N & \bm a_2\bm 1^T &\bm I_N - \frac{1}{N}\bm 1\bm 1^T\\\bm 0_N & \bm 0_N &\bm I_N \end{bmatrix} $$
for fixed $\bm a_1, \bm a_2\in\mathbb{R}^N$.
\end{theorem}
\begin{IEEEproof}
By Lemma \ref{lem:linearconvergence_DCADMM}, we have for any $\bm s^0$ that satisfies the initialization condition, $$\bm s^\infty = \left[\begin{array}{c} \bm x^{\infty} \\\bm\alpha^{\infty}\\\bm r^{\infty} \end{array} \right] = \left[\begin{array}{c} \bm x^{*} \\\bm\alpha^{*}\\\bm r^{*} \end{array} \right]=\left[\begin{array}{c} \bm1x_\text{avg}\\\bm r - \bm 1x_\text{avg}\\\bm r \end{array} \right].$$Recall that $\bm s^\infty=\bm D^\infty \bm s^0$. If we fix $\bm\alpha^0$ and $\bm r^0$, global convergence implies that $\bm s^\infty=\bm s^*$ regardless of the initial value $\bm x^0$. Thus $\bm D_{i1}=\bm 0_N$, $i=1,2,3$. Similarly, fixing $\bm x^0$ and $\bm r^0$, we must have $\bm D_{12}\bm\alpha^0=\bm D_{22}\bm \alpha^0=\bm 0$. Since $\bm\alpha^0$ is initialized in the column space of $\bm M_-\bm M_-^T=2\bm L_-$ where $\bm L_-$ is the signed Laplacian matrix of a connected undirected graph, $\bm D_{12}$ and $\bm D_{22}$ must be respectively the products of some vectors $\bm a_1$ and $\bm a_2$ in $\mathbb{R}^N$ multiplying $\bm 1^T$, such that $\bm D_{12}\bm L_-=\bm D_{22}\bm L_-=\bm 0$. Knowing the form of $\bm D_{j1}$ and $\bm D_{j2}$,~$j=1,2$, we see that $\bm x^\infty$ and $\bm\alpha^\infty$ only depend on $\bm r^0=\bm r$. Together with the facts that $\bm x^\infty=\bm x^*$ has each entry of itself reaching the data average $x_\text{avg}=\frac{1}{N}\bm r^T \bm 1$ and that $\bm\alpha^\infty=\bm r - \bm 1x_\text{avg}$ for any $\bm r$, we validate $\bm D_{13}$ and $\bm D_{23}$ as given in the theorem. The remaining blocks, $\bm D_{32}$ and $\bm D_{33}$, follow directly from the matrix multiplication.
\end{IEEEproof}

Given global convergence, we now turn our attention to the rate of convergence of the CADMM. Recent work of \cite{Hong2012, Deng2016}
has established the linear convergence of the ADMM. Unfortunately, their results do not apply to the CADMM as their conditions are not satisfied here. In \cite{Hong2012}, the step size of the dual variable update, i.e., $\rho$ in the $\bm\lambda$-update of (\ref{eqn:admmupdates}), need be sufficiently small while our CADMM has a fixed step size $\rho$ that can be any positive number (see Remark \ref{rmk:rho} for further discussion on the choice of $\rho$). The linear convergence in \cite{Deng2016} is established provided that either $g(\bm z)$ is strongly convex or $\bm B$ is full row-rank in (\ref{eqn:admmformulation}). In our formulation, however, $g(\bm z)=0$ is not strongly convex and $\bm B=[-\bm I_{2E};-\bm I_{2E}]$ is row-rank deficient. Nevertheless, we first give Lemma \ref{lem:linearconvergence} with regards to the convergence rate of a vector concatenating $\bm z$ and $\bm \beta$. A more general result can be found in \cite[Theorem 1]{Shi2014}. Our proof is similar to that of \cite{Shi2014} but simpler.

\begin{lemma}[{\cite[Theorem 1]{Shi2014}}]
\label{lem:linearconvergence} 
Consider the ADMM iteration (\ref{eqn:admmupdates}) that solves (\ref{eqn:matrixform}). Define $$\bm u = \left[\begin{array}{c} \bm z\\\bm \beta \end{array} \right]~\text{and}~\bm G = \begin{bmatrix} \rho \bm I_{2E} & \bm 0_{2E} \\ \bm 0_{2E}& \frac{1}{\rho}\bm I_{2E}  \end{bmatrix},$$where $\bm\beta$ is the dual variable. If we initialize $\bm z^0=\frac{1}{2}\bm M_+^T\bm x^0$, $\bm \beta^0=-\bm\gamma^0$ where $\bm\gamma$ is the other dual variable and $\bm\beta^0$ is in the column space of $\bm M_-^T$, then for $k=0,1,\cdots$, $\bm z^k=\frac{1}{2}\bm M_+^T\bm x^k$, $\bm\beta^k$ lies in the column space of $\bm M_-^T$, and $(\bm x^k,\bm z^k,\bm\beta^k)$ converges uniquely to $(\bm x^*,\bm z^*,\bm \beta^*)$ with $\bm x^*=\bm 1x_\text{avg}$, $\bm z^*=\frac{1}{2}\bm M_+^T\bm 1x_{\text{avg}}$ and $\bm\beta^*$ being a vector in the column space of $\bm M_-^T$. Furthermore, $\bm u^k = [\bm z^k;\bm\beta^k]$ converges Q-linearly to its optimal $\bm u^*=[\bm z^*;\bm\beta^*]$ with respect to the $\bm G$-norm
\begin{equation}
\label{eqn:Ulinearconvergence}
\begin{aligned}
\|\bm u^{k+1}-\bm u^*\|_{\bm G}^2\leq\frac{1}{1+\delta}\|\bm u^k-\bm u^*\|_{\bm G}^2,
\end{aligned}
\end{equation}
where $$\delta = \min\left\{\frac{{\tilde{\sigma}}_{\min}^2(\bm M_-)}{2\sigma_{\max}^2(\bm M_+)}, \frac{4\rho{\tilde{\sigma}}_{\min}^2(\bm M_-)}{\rho^2\sigma_{\max}^2(\bm M_+){\tilde{\sigma}}_{\min}^2(\bm M_-)+ 8}\right\},$$
${\sigma_{\max}}(\bm M_+)$ denotes the spectral norm or the largest singular value of $\bm M_+$, and $\tilde{\sigma}_{\min}(\bm M_-)$ denotes the smallest positive singular value of $\bm M_-$.
\end{lemma}
\begin{IEEEproof}
See Appendix.
\end{IEEEproof}

With this lemma, we can now establish the linear convergence rate of the CADMM .

\begin{theorem}[Linear convergence of the CADMM]
\label{thm:linearconandbound}
Consider the matrix form of the CADMM in (\ref{eqn:iterateform}). If $\bm s^0=[\bm x^0;\bm\alpha^0;\bm r^0]$ satisfy the initialization condition, then 
\begin{align}
\|\bm s^{k+1}-\bm s^*\|_2 &=\|(\bm D^{k+1}-\bm D^*)\bm s^0\|_2\nn\\&\leq \left(1+\sqrt{\frac{\rho}{1+\delta}}\sigma_{\max}(\bm M_-)\right)\|\bm u^k-\bm u^*\|_{\bm G},\nn
\end{align} where $\delta$ and $\bm u^k$ are defined in Lemma \ref{lem:linearconvergence}. Therefore, $\bm s^k$ is R-linearly convergent to $\bm s^*$.
\end{theorem}
\begin{IEEEproof}
Notice that the initializations in Lemma \ref{lem:linearconvergence} decentralize the ADMM iteration (\ref{eqn:admmupdates}) into the CADMM. Thus $\bm x^k$ is the same in the ADMM iteration (\ref{eqn:admmupdates}) and the CADMM iteration (\ref{eqn:distributedversiond_plug}) while $\bm\alpha^k=\bm M_-\bm\beta^k$. Then (\ref{eqn:x13}) implies 
$$\|\bm x^{k+1}-\bm x^*\|_2\leq\|\bm u^k-\bm u^*\|_{\bm G}.$$ 
We also have  
\begin{align}
\|\bm \alpha^{k+1}-\bm\alpha^*\|_2&=\|\bm M_-(\bm\beta^{k+1}-\bm\beta^*)\|_2\nn\\&\leq\sigma_{\max}(\bm M_-)\|\bm\beta^{k+1}-\bm\beta^*\|_2\nn\\\label{eqn:alphau}&\leq\sqrt{\rho}\sigma_{\max}(\bm M_-)\|\bm u^{k+1}-\bm u^*\|_{\bm G}\\&\leq\sqrt{\frac{\rho}{1+\delta}}\sigma_{\max}(\bm M_-)\|\bm u^{k}-\bm u^*\|_{\bm G},\nn
\end{align}
where the last two inequalities are from the definitions of $\bm u$ and $\bm G$, and (\ref{eqn:Ulinearconvergence}), respectively.
Thus, 
\begin{align}
\|\bm s^{k+1}-\bm s^*\|_2&\leq\|\bm x^{k+1}-\bm x^*\|_2+\|\bm \alpha^{k+1}-\bm\alpha^*\|_2\nn\\
&\leq\left(1+\sqrt{\frac{\rho}{1+\delta}}\sigma_{\max}(\bm M_-)\right)\|\bm u^k-\bm u^*\|_{\bm G}.\nn
\end{align}
\end{IEEEproof}
\section{Quantized Consensus}
\label{sec:problemformulation}
To model the effect of quantized communications, we assume that each agent can store and compute real values with infinite precision; however, an agent can only transmit quantized data through the channel which are received by its neighbors without any error. The quantization operation is defined as follows. Let $\Delta>0$ be a given quantization resolution and define the quantization lattice in $\mathbb{R}$ by $$ \Lambda = \{t\Delta: t\in\mathbb{Z}\}.$$ A quantizer is a function $Q: \mathbb{R}\to\Lambda$ that maps a real value to some point in $\Lambda$. 
Among all quantizers we consider the following two for distributed average consensus:
\begin{enumerate}
\item Probabilistic quantizer $Q_p$ defined as follows:
for $y \in\left [t\Delta,(t+1)\Delta\right)$,
\begin{equation}
\label{eqn:Qp}
\begin{aligned}
\hspace{-0.08in}Q_p(y) =
\begin{cases}
t\Delta,&\hspace{-0.05in}\text{with probability}~t+1-\frac{y}{\Delta}, \\
(t+1)\Delta,&\hspace{-0.05in}\text{with probability}~\frac{y}{\Delta}-t.
\end{cases}
\end{aligned}
\end{equation}\item Rounding quantizer $Q_d$ which projects $y\in\mathbb{R}$ to its nearest point in $\Lambda$:
\begin{equation}
\label{eqn:Qd}
\begin{aligned}
Q_d(y) = t \Delta,~\text{if}~\left(t-\frac{1}{2}\right)\Delta\leq y< \left(t+\frac{1}{2}\right)\Delta.
\end{aligned}
\end{equation}\end{enumerate}
We point out that probabilistic quantization is equivalent to a dithered quantization method (see \cite[Lemma 2]{Aysal2008}) while rounding quantization is one of the deterministic quantization schemes. Throughout the rest of this paper, we use $Q(y)$ (or $y_{[Q]}$ for ease of presentation) to denote the quantized value of $y\in\mathbb{R}$ regardless of its quantization scheme; we use $Q_p(y)$ (or $y_{[Q_p]}$) and $Q_d(y)$ (or $y_{[Q_d]}$) when it is necessary to specify the quantization scheme. Quantizing a vector means quantizing each of its entries. Define $e_{Q}=y_{[Q]}-y$ as the quantization error. It is clear that
\begin{equation}
\label{eqn:Qerror}
\begin{aligned}
\left |e_{Q_p}\right|\leq\Delta~\text{and}~\left|e_{Q_d}\right|\leq \frac{1}{2}\Delta,~~\text{for any}~y\in\mathbb{R}.
\end{aligned}
\end{equation}


As seen from Section \ref{sec:ADMMnoQ}, the CADMM has the advantage of global and linear convergence for solving the average consensus problem as long as the initialization condition is met.  The authors in \cite{Zhu2009, Erseghe2011} have also shown the good behavior of the ADMM in distributed settings when noise or random link failures are imposed. The rest of this paper is devoted to investigating the effects of the two quantization schemes defined in (\ref{eqn:Qp}) and (\ref{eqn:Qd}) on the performance of the CADMM. We remark that the results of probabilistic and rounding quantizations hold respectively for other dithered and deterministic cases, which will be elaborated in Sections \ref{sec:PQ} and \ref{sec:DQ}. 


\section{Probabilistic Quantization}
\label{sec:PQ}

For ease of presentation, we only study the probabilistic quantization defined in (\ref{eqn:Qp}). The results can be easily extended to any other dithered quantization as the only information used is the first and second order moments of the probabilistic quantizer output which are stated in the following lemma. See \cite{Xiao2005a} for a proof.

\begin{lemma}[{\cite[Lemma 2]{Xiao2005a}}]
\label{lem:QPproperty}
For every $y\in\mathbb{R}$, it holds that 
\begin{equation}
\begin{aligned}
\mathbb{E}\left[Q_p(y)\right]=y~\text{and}~\mathbb{E}\left[\left(y-Q_p(y)\right)^2\right]\leq\frac{\Delta^2}{4}.\nn
\end{aligned}
\end{equation}
\end{lemma}
The iteration in (\ref{eqn:distributedversiond_plug}) now takes the form of
\begin{equation}
\label{eqn:distributedversiond_plug_Qp}
\begin{aligned}
&x_i^{k+1}=\frac{1}{1+2\rho|\mathcal{N}_i|}\Bigg(\rho|\mathcal{N}_i|x_{i[Q_p]}^k+\rho\sum_{j\in\mathcal{N}_i} x_{j[Q_p]}^k-\alpha_i^k+r_i\Bigg),\\
&\alpha_i^{k+1}=\alpha_i^k+\rho\Bigg(|\mathcal{N}_i|x_{i[Q_p]}^{k+1}-\sum_{j\in\mathcal{N}_i} x_{j[Q_p]}^{k+1}\Bigg).
\end{aligned}
\end{equation} 
Notice that $x_i^k$ is also quantized at its own node for the $(k+1)$th update; the reason will be given in Remark~\ref{rmk:Qitself}. As illustrated in \cite{Zhu2009}, iteration (\ref{eqn:distributedversiond_plug_Qp}) can be interpreted as a stochastic gradient update. Viewed from this point, the quantization error causes $x_i^{k}$ to fluctuate around the quantization-free updates  (\ref{eqn:distributedversiond_plug}). Our convergence claims are given in Theorem \ref{thm:Qpresult}.
\begin{theorem}
\label{thm:Qpresult}
Let $\bm x^0$ and  $\bm\alpha^0$ satisfy the initialization condition. The probabilistically quantized CADMM (PQ-CADMM) iteration (\ref{eqn:distributedversiond_plug_Qp}) generates $x_i^k,i=1,2,\cdots,N$, which converges linearly to the data average $x_\text{avg}$ in the mean sense as $k\to\infty$. In addition, the variance of $x_i^k$ converges to a finite value which depends on $\Delta$ and the network topology.
\end{theorem}
\begin{IEEEproof}
Taking expectation of both sides of  (\ref{eqn:distributedversiond_plug_Qp}), we have 
\begin{equation}
\label{eqn:distributedversiond_plug_Qp_Expectation}
\begin{aligned}
&\mathbb{E}[x_i^{k+1}]=\frac{1}{1+2\rho|\mathcal{N}_i|}\Bigg(\rho|\mathcal{N}_i| \mathbb{E}[x_{i[Q_p]}^k]+\rho\sum_{j\in\mathcal{N}_i} \mathbb{E}[x_{j[Q_p]}^k]\\&\hspace{2.2in}-\mathbb{E}[\alpha_i^k]+r_i\Bigg),\\
& \mathbb{E}[\alpha_i^{k+1}]= \mathbb{E}[\alpha_i^k]+\rho\Bigg(|\mathcal{N}_i| \mathbb{E}[x_{i[Q_p]}^{k+1}]-\sum_{j\in\mathcal{N}_i}  \mathbb{E}[x_{j[Q_p]}^{k+1}]\Bigg).
\end{aligned}
\end{equation} 
Noting that Lemma \ref{lem:QPproperty} implies $\mathbb{E}[x_{i[Q_p]}^k]=\mathbb{E}[x_i^k]$ and $\mathbb{E}[x_{j[Q_p]}^k]=\mathbb{E}[x_j^k]$, we see that (\ref{eqn:distributedversiond_plug_Qp_Expectation}) takes exactly the same iterations in the mean sense as the CADMM. By initializing $\bm\alpha^0$ in the column space of $\bm L_-$, $\mathbb{E}[\bm\alpha^0]=\bm\alpha_0$ satisfies the initialization condition. The linear convergence of $\mathbb{E}[x_i^k]$ to $x_\text{avg}$ is thus ensured due to Theorem \ref{thm:linearconandbound}.

Since Lemma \ref{lem:QPproperty} also indicates the bounded variance of quantization error, the second claim follows directly from \cite[Proposition 3]{Zhu2009}.
\end{IEEEproof}

We notice that the convergence of $\mathbb{E}[x_i^k]\to x_\text{avg}$ does not indicate that $\bm x^k$ reaches a consensus when $k\to\infty$. Nevertheless, a simple method fixes this problem. The idea is to calculate the running average $\bar{x}_i^k=\frac{1}{k}\sum_{l=1}^k x_i^l,k\geq1$ at each node $i$. One can use similar steps in the proof of \cite[Proposition 3]{Zhu2009} to show that $\bar{x}_i^k$ has diminishing variance. By Chebyshev's inequality, we then get the following corollary.
\begin{corollary}
Let $\bar{x}_i^k=\frac{1}{k}\sum_{l=1}^k x_i^l$ for $k\geq1$. For each node $i$, we have $$\mathbb{P}\left[\lim_{k\to\infty}\bar{x}_i^k=x_\text{avg}\right]=1.$$
\end{corollary}
 
\section{Deterministic Quantization} 
\label{sec:DQ}
Deterministic quantization is usually much harder to handle as the quantization error is not stochastic. Unlike probabilistic quantization, the accumulated error term is very likely to blow up; there have been a few methods proposed to counter such difficulties (see \cite{Nedic2009,Chamie2014,Carli2010}), yet the resulting algorithms either do not guarantee a consensus or reach a consensus with an error from the desired average that depends on the number of agents, the quantization resolution, and the agents' data. Our approach will establish a finite upper bound on the accumulated error term and then use the property and the initialization condition of local Lagrangian multipliers to deduce the consensus reaching result.

Let the local data $x_i^k$ be also quantized for the $(k+1)$th update at node $i$. The updates become
\begin{equation}
\label{eqn:Qdistributedversion}
\begin{aligned}
&x_i^{k+1}=\frac{1}{1+2\rho|\mathcal{N}_i|}\Bigg(\rho|\mathcal{N}_i|x_{i[Q_d]}^k+\rho\sum_{j\in\mathcal{N}_i} x_{j[Q_d]}^k-\alpha_i^k+r_i\Bigg),\\
&\alpha_i^{k+1}=\alpha_i^k+\rho\Bigg(|\mathcal{N}_i|x_{i[Q_d]}^{k+1}-\sum_{j\in\mathcal{N}_i} x_{j[Q_d]}^{k+1}\Bigg).
\end{aligned}
\end{equation}
Rewrite $x_{i[Q_d]}^k=x_i^k+e_{iQ_d}^k$ with $e_{iQ_d}^k\in[-\Delta/2,\Delta/2)$ according to (\ref{eqn:Qd}). Then the $\alpha_i$-update, $i=1,2,\cdots,N$, is equivalent to 
\begin{align}
\alpha_i^{k+1} = \alpha^k_i&+\rho\Bigg(|\mathcal{N}_i|x_{i}^{k+1}-\sum_{j\in\mathcal{N}_i} x_{j}^{k+1}\Bigg)\nn\\
&~~~~~~~~~~~~~~~+\rho\Bigg(|\mathcal{N}_i|e_{iQ_d}^{k+1}-\sum_{j\in\mathcal{N}_i} e_{jQ_d}^{k+1}\Bigg),\nn
\end{align}
or written in the matrix form,
\begin{align}
\label{eqn:alphamatrixupdate}
\bm\alpha^{k+1}=\bm\alpha^k+\rho{\bm L}_-\bm x^{k+1}+\rho\bm L_-\bm e^{k+1}_{Q_d},
\end{align}
where $\bm e_{Q_d}^k$ denotes the vector concatenating all $e_{iQ_d}^k$. Recalling the ideal CADMM update (\ref{eqn:idealdcadmm}), we have the matrix form of (\ref{eqn:Qdistributedversion}) as
\begin{equation}
\label{eqn:iterateform_QD}
\begin{aligned}
\bm s^{k+1}=\bm D(\bm s^k+\bm {s}_{x}^{k})+\bm {s}_\alpha^k
\end{aligned}
\end{equation}
where $\bm s_x^k=[\bm e^k_{Q_d};\bm 0;\bm 0]$ and $\bm s_\alpha^k=[\bm 0; \rho\bm L_- \bm e^{k+1}_{Q_d};\bm 0]$. It is important to note that $\bm e_{Qd}^k$ is deterministic and hence the update (\ref{eqn:iterateform_QD}) is deterministic. Our main results are stated in the following theorem.
\begin{theorem}
\label{thm:mainresults}
Consider the deterministically quantized CADMM (DQ-CADMM) iteration (\ref{eqn:Qdistributedversion}). Let $\bm x^0$ and $\bm\alpha^0$ satisfy the initialization condition for the CADMM. Then there exists a finite time iteration $k_0\geq1$ such that for $k\geq k_0$ all the quantized variable values
\begin{itemize}
\item either converge to the same quantization value:
$$x_{1[Qd]}^k=\cdots=x_{N[Qd]}^k\triangleq x_{[Qd]}^*,$$

\item or cycle around the average $x_\text{avg}$ with a finite period $T\geq2$, i.e., $x_{i[Qd]}^k=x_{i[Qd]}^{k+T}, i=1,2,\cdots, N$, and 
\begin{align}
\label{eqn:cyclic_consensus}
\frac{1}{T}\hspace{-1pt}\sum_{l=1}^{T}x_{1[Qd]}^{k+l}=\cdots =\frac{1}{T}\hspace{-1pt}\sum_{l=1}^{T}x_{N[Qd]}^{k+l}\triangleq\bar{x}_{[Qd]}^{*}.
\end{align}
\end{itemize}

For both convergent and cyclic cases, we have the following error bound for $x^*_{Qd}\in\{x_{[Qd]}^*, \bar{x}_{[Qd]}^*\}$:
\begin{align}
\label{eqn:consensuserror}
\left|x_{Qd}^*-x_\text{avg}\right|\leq \left(\frac{1}{2}+\rho\frac{2E}{N}\right)\Delta,
\end{align}
where the upper bound is tight if the DQ-CADMM converges.


\end{theorem}
\begin{IEEEproof} We prove that the DQ-CADMM either converges or cycles after a finite-time iteration and then use this fact to derive the error bounds.

We see from (\ref{eqn:alphamatrixupdate}) that $\bm\alpha^k$ must lie in the column space of $\bm L_-$ if $\bm\alpha^0$ is initialized in the column space of $\bm L_-$. Following (\ref{eqn:iterateform_QD}), we have 
\begin{align}
\bm s^{k} &= \bm D(\bm s^{k-1}+ \bm s_x^{k-1})+\bm s_\alpha^{k-1}\nn\\
&=\bm D\left(\bm D(\bm s^{k-2}+\bm s_x^{k-2})+\bm s_\alpha^{k-2}\right)+\bm D \bm s_x^{k-1} + \bm s_\alpha^{k-1}\nn\\
&=\cdots\nn\\
\label{eqn:expansion}
&=\bm D^k \bm s^0+\left(\sum_{i=1}^{k} \bm D^i \bm s_x^{k-i}+\sum_{j=0}^{k-1} \bm D^j\bm s_\alpha^{k-1-j}\right).
\end{align}
The first term is simply the ideal CADMM update which converges to a finite value. We will show that the accumulated error term $\sum_{i=1}^{k} \bm D^i \bm s_x^{k-i}+\sum_{j=0}^{k-1} \bm D^j\bm s_\alpha^{k-1-j}$ is bounded and hence that $\bm s^k$ is bounded. Notice that $\bm D^i\bm s_x^{k-i}$ is the $i$th update of the CADMM with the initial value $\bm s_x^{k-i}$. Let $\bm u_{k-i}^l=[\bm z_{k-i}^l;\bm\beta_{k-i}^l]$ be the vector that concatenates the primal and dual variables in the ADMM iteration (\ref{eqn:admmupdates}), with initial values $\bm z_{k-i}^0= \frac{1}{2}\bm M_+^T\bm e^{k-i}_{Q_d}$ and $\bm\beta_{k-i}^0=\bm 0$ corresponding to $\bm s_x^k=[\bm e^k_{Q_d};\bm 0;\bm 0]$. With $\bm G$ defined in Lemma \ref{lem:linearconvergence}, we obtain
\begin{align}
\|\bm u_{k-i}^0\|^2_{\bm G}&=\rho\left\|\frac{1}{2}\bm M_+^T\bm e_{Q_d}^{k-i}\right\|_2^2\nn\\
&\leq\frac{1}{4}\rho\sigma_{\max}^2(\bm M_+)\|\bm e_{Q_d}^{k-i}\|_2^2\nn\\
&\leq \frac{1}{16}\rho N\Delta^2\sigma_{\max}^2(\bm M_+),\nn
\end{align}
where the last inequality is from (\ref{eqn:Qerror}).
Since Theorem \ref{thm:Dproperty} indicates the form of $\bm D^*$, we get $\bm D^* \bm s_x^{k-i}=\bm 0$, i.e., $\bm x_{k-i}^*=\bm 0$ and $\bm\alpha_{k-i}^*=\bm 0$. Therefore, $\bm u_{k-i}^*=[\bm z_{k-i}^*; \bm\beta_{k-i}^*]=\bm 0$ from Lemma \ref{lem:abrelation} and the fact that $\bm z_{k-i}^*=\frac{1}{2}\bm M_+^T\bm x_{k-i}^*$. Noting also that the initialization $\bm z_{k-i}^0$ and $\bm \beta_{k-i}^0$ meet the condition of Lemma \ref{lem:linearconvergence}, we thus have 
\begin{align}
\label{eqn:sx}
\|\bm D^i \bm s_x^{k-i}\|_2 &= \|(\bm D^{i} -\bm D^*)\bm s_x^{k-i}\|_2\nn\\
&\stackrel{(a)}{\leq}\left(1+\sqrt{\frac{\rho}{1+\delta}}\sigma_{\max}(\bm M_-)\right)\|\bm u_{k-i}^{i-1}-\bm u_{k-i}^*\|_{\bm G}\nn\\
&\stackrel{(b)}{\leq}\left(1+\sqrt{\frac{\rho}{1+\delta}}\sigma_{\max}(\bm M_-)\right) \left(\sqrt{\frac{1}{1+\delta}}\right)^{i-1}\nn\\&~~~~\times\frac{1}{4}\Delta\sigma_{\max}(\bm M_+)\sqrt{\rho N},
\end{align}
where $(a)$ is from Theorem \ref{thm:linearconandbound} and $(b)$ is due to Lemma \ref{lem:linearconvergence} together with the fact that $\bm u_{k-i}^*=\bm 0$. Similarly, we have for $j\geq 1$,
\begin{align}
\label{eqn:sa}
\|\bm D^j \bm s_\alpha^{k-1-j}\|_2 &\leq \left(1+\sqrt{\frac{\rho}{1+\delta}}\sigma_{\max}(\bm M_-)\right) \left(\sqrt{\frac{1}{1+\delta}}\right)^{j-1}\nn\\&~~~~\times\frac{1}{4}\Delta\sigma_{\max}(\bm M_-)\sqrt{{\rho}N},
\end{align}
and when $j=0$, 
\begin{align}
\label{eqn:sa0}
\|\bm D^j \bm s_\alpha^{k-1-j}\|_2=\|\bm s_\alpha^{k-1}\|_2\leq \frac{1}{4}\Delta\sigma_{\max}(\bm M_-)\rho\sqrt{N}.
\end{align} 
Therefore, 
\begin{align}
&~~~~\left\|\sum_{i=1}^{k} \bm D^i \bm s_x^{k-i}+\sum_{j=0}^{k-1} \bm D^j\bm s_\alpha^{k-1-j}\right\|_2\nn\\
&\leq \sum_{i=1}^{k} \|\bm D^i \bm s_e^{k-i}\|_2+\sum_{j=0}^{k-1} \|\bm D^j\bm s_\alpha^{k-1-j}\|_2\nn\\
&\leq \|\bm s_\alpha^{k-1}\|_2+\sum_{i=1}^{k}\left(\|\bm D^i \bm s_e^{k-i}\|_2+ \|\bm D^i\bm s_\alpha^{k-1-i}\|_2\right)\nn\\
&\stackrel{(a)}{\leq}\frac{1}{4}\Delta\sigma_{\max}(\bm M_-)\rho\sqrt{N}+\left(1+\sqrt{\frac{\rho}{1+\delta}}\sigma_{\max}(\bm M_-)\right)\nn\\       
&~~~~\times\frac{1}{4}\Delta\sqrt{\rho N}\left(\sigma_{\max}(\bm M_+)+\sigma_{\max}(\bm M_-)\right)\sum_{i=1}^{k}\left(\sqrt{\frac{1}{1+\delta}}\right)^{i-1}
\label{eqn:ddd}
\end{align}
where $(a)$ is from (\ref{eqn:sx})-(\ref{eqn:sa0}). Then (\ref{eqn:ddd}) must be finite for $k=1,2,\cdots,$ as $\delta >0$, and thus $\bm s^k$ is bounded. An important fact from (\ref{eqn:iterateform_QD}) is that the update of $\bm s^{k+1}$ and hence $\bm s_x^{k+1}$ is fully determined by $\bm s^k+\bm s_x^k$ due to the deterministic quantization and the CADMM update. Recalling that $\|\bm s_x^k\|_2 =\|\bm e_{Qd}^k\|_2\leq\frac{\Delta}{2}\sqrt{N}$ and that $\bm s^k+\bm s_x^k = [\bm x^k_{[Qd]};\bm\alpha^k;\bm r]$ with each entry of $\bm x^k_{[Qd]}$ being a multiple of $\Delta$, each entry of $\bm\alpha$ being a multiple of $\rho\Delta$, and $\bm r$ being fixed, we conclude that there are only finite possible states of $\bm s^k+\bm s_x^k$. Therefore, $\bm s^k$ is either convergent or cyclic with a finite period $T\geq 2$ after a finite-time iteration. 

We next consider error bounds for the consensus value. The consensus error may be studied directly by calculating the accumulated error term in (\ref{eqn:expansion}). However, the bound in (\ref{eqn:ddd}) is quite loose in general since it results from the worst case. We alternatively derive the error bounds in the respective case using the fact that the DQ-CADMM either converges or cycles.

\emph{Convergent case:} The convergence of the DQ-CADMM implies that $\bm s^{k+1}=\bm s^k$ for $k\geq k_0$, and hence 
$$\bm 0 = \bm\alpha^{k+1}-\bm\alpha^k=\rho\bm L_-\bm x^{k+1}_{[Qd]}.$$
Since $\bm L_-$ is the Laplacian matrix of a connected graph $\mathcal{G}_u$, we must have that $\bm x^{k+1}_{[Qd]}$ reaches a consensus. Now let $x_{[Q_d]}^*\in\Lambda$ denote the convergent quantized value. Then $x_{i[Q_d]}^\infty=x_{[Q_d]}^*$ for $i = 1,2,\cdots,N$, and $x_i^\infty=x_{[Q_d]}^*-e_{iQ_d}^*$. Summing up both sides of (\ref{eqn:Qdistributedversion}) from $i = 1$ to $N$, we have 
\begin{align}
\sum_{i=1}^N (1+2\rho|\mathcal{N}_i|)&\left(x_{[Q_d]}^*-e_{iQ_d}^*\right)\nn\\&= \sum_{i=1}^N\bigg(\rho|\mathcal{N}_i| x_{[Q_d]}^*+\rho\sum_{j\in\mathcal{N}_i} x_{[Q_d]}^*+r_i\bigg),\nn
\end{align}
which is equivalent to 
$$x^*_{[Q_d]} = \frac{1}{N}\sum_{i=1}^N r_i + \frac{1}{N}\sum_{i=1}^N (1+2\rho|\mathcal{N}_i|) e_{iQ_d}^*.$$
Here we use the fact that $\bm \alpha^k$ lies in the column space of $\bm L_-$, i.e.,  $\bm \alpha^k=\bm L_-\bm b^k$ where $\bm b^k\in\mathbb{R}^N$. Then $\sum_{i=1}^N \alpha^k_i = (\bm L_-\bm b^k)^T\bm 1=(\bm b^k)^T(\bm L_-^T\bm 1)=0$.
Recalling that $x_\text{avg} = \frac{1}{N}\sum_{i=1}^N r_i$ and $|e_{iQ_d}|\leq\frac{\Delta}{2}$, we finally obtain
\begin{align}
\left|x^*_{[Q_d]}-x_\text{avg}\right|\leq \left(\frac{1}{2}+\rho\frac{2E}{N}\right)\Delta.\nn
\end{align}

The following example shows the tightness of this bound in this convergent case. Consider a simple two-node network with $r_1=-\frac{3}{2}$ and $ r_2=-\frac{7}{2}$. Set both $\Delta$ and $\rho$ to be $1$. In this case, we have $E=1$, $N=2$ and $$\bm L_-=\begin{bmatrix} 1& -1\\ -1& 1\end{bmatrix}.$$ We start with $x^0_{1[Q_d]}=x^0_{2[Q_d]}=-1$ and $\alpha_1^0= -\alpha_2^0=1$. One can easily check that our initialization condition is met, and $x^k_{1[Q_d]}= x^k_{2[Q_d]}=-1$ and $\alpha_1^k= -\alpha_2^k=1, k = 1,2,\cdots$, in the updates of (\ref{eqn:Qdistributedversion}). Hence $x^*_{[Q_d]}=-1$ and the consensus error is 
\begin{align}
\left |x_{[Q_d]}^*-x_\text{avg}\right |=\frac{3}{2}=\left(\frac{1}{2}+\rho\frac{2E}{N}\right)\Delta.\nn
\end{align}
This coincides with the error bound in (\ref{eqn:consensuserror}).

\emph{Cyclic case:} When the DQ-CADMM cycles with a period $T$, we must have $\bm s^{k+T}=\bm s^k$. Thus, for $k\geq k_0$, we have that 
$$\bm 0 = \bm\alpha^{k+T}-\bm\alpha^k = \rho\bm L_-\sum_{l=1}^{T}\bm x^{k+l}_{[Qd]},$$
and consequently, $\sum_{l=1}^{T}\bm x^{k+l}_{[Qd]}$ reaches a consensus, i.e., (\ref{eqn:cyclic_consensus}) is true. Now denote $$\bar{x}^*_{[Qd]} = \frac{1}{T}\sum_{l=1}^{T}x^{k+l}_{i[Qd]}, i=1,\cdots, n.$$ We then get
\begin{align}
\label{eqn:cycbd}\left|\bar{x}_{[Qd]}^*-\frac{1}{T}\sum_{l=1}^{T}x^{k+l}_{i}\right|\leq \frac{1}{T}\sum_{l=1}^T\left|x^{k+l}_{i[Qd]}-x^{k+l}_i\right|\leq\frac{\Delta}{2}.
\end{align}
Summing both sides of (\ref{eqn:Qdistributedversion}) over one period and dividing the sum by $T$, we have 
$$\frac{1}{T}\sum_{l=1}^T x_i^{k+l}\hspace{-2pt}=\hspace{-2pt}\frac{1}{1+2\rho|\mathcal{N}_i|}\hspace{-2pt}\left(\hspace{-2pt}2\rho|\mathcal{N}_i|\bar{x}^*_{[Qd]}-\frac{1}{T}\sum_{l=1}^T\alpha_i^{k+l}+r_i\hspace{-2pt}\right)\hspace{-1pt}.$$
Finally, using (\ref{eqn:cycbd}) and following the same steps as in the convergent case we conclude that 
$$\left|\bar{x}^*_{[Qd]}-x_\text{avg}\right|\leq \left(\frac{1}{2}+\rho\frac{2E}{N}\right)\Delta.$$
\end{IEEEproof}
\begin{remark}
The result that deterministic quantization may lead the consensus algorithm to either convergent or cyclic cases is also reported in \cite{Chamie2014}. Similar to theirs, one can use the history of agents' variables, e.g., running average, to achieve asymptotic convergence at each node. Differently, while they can make local variable values close to the true average in cyclic cases without guaranteeing  a consensus, our algorithm can reach a consensus but does not make the error arbitrarily small in general. 
\end{remark}
\begin{remark}
\label{rmk:notglobal}
We shall mention that $x^*_{[Qd]}$ or $\bar{x}^*_{[Qd]}$ need not be unique. This is because, unlike the ideal CADMM, $\|\bm u^k - \bm u^*\|_{\bm G}$ in the DQ-CADMM need not decrease monotonically due to the quantization {\em that occurs on $\bm x^k$ at each update}. Note also that practical consensus value does not necessarily meet the error bound and we usually have smaller errors than (\ref{eqn:consensuserror}) in practice (see Fig.~\ref{fig:cerr}). We hence expect better consensuses when $(\bm x^0,\bm\alpha^0)$ are initialized closer to the ideal optima, which leads to a two-stage algorithm for quantized consensus in Section~\ref{sec:algorithm}.
\end{remark}
\begin{remark} 
\label{rmk:rho}
An interesting observation of our main result is the ADMM parameter $\rho$. While a small $\rho$ indicates a small consensus error bound, the current paper does not quantify how it affects the convergence rate. Here we do not study the optimal selection of $\rho$ but simply set $\rho=1$. Therefore we do not regard $\rho$ as a factor affecting our algorithm's performance. We refer readers to \cite{BoydADMM,Shi2014,Ghadimi2015ParaSel} for detailed discussions on how $\rho$ affects the ADMM's performance.  
\end{remark}
\begin{remark}
Theorem~\ref{thm:mainresults} for rounding quantization extends straightforward to other deterministic quantizations as the only information used in our proof is the bounded quantization error. In contrast with \cite{Kashyap2007,Nedic2009} where the algorithms may fail for some deterministic quantization schemes, e.g., the rounding quantization, our results work for all deterministic quantization schemes as long as a finite quantization error bound is provided. 
\end{remark}
\begin{remark}
\label{rmk:Qitself}
In both the PQ-CADMM and DQ-CADMM iterations, $x^k_i$ is quantized for the $(k+1)$th update at node $i$ even though nodes can compute and store real values with infinite precision. The reason is to guarantee that $\bm\alpha^k$ lies in the column space of $\bm L_-$ and thus the ideal CADMM update in either the PQ-CADMM or the DQ-CADMM [cf. Equation (\ref{eqn:iterateform_QD})] possesses the linear convergence property given in Theorem \ref{thm:linearconandbound}. If we do not quantize $x^k_i$ at its own node, Theorem \ref{thm:Qpresult} still holds due to $\mathbb{E}[x^k_{i[Q_p]}] = \mathbb{E}[x_i^k]$ while Theorem \ref{thm:mainresults} may fail.
\end{remark}
\begin{remark}
In the problem reformulation (\ref{eqn:admmformulation}), each node $i$ has its local objective function being $\frac{1}{2}(x_i-r_i)^2$ and $\bm1x_\text{avg}$ minimizes the global objective function $f(\bm x)$ which is the sum of the local objectives. To analyze the DQ-CADMM, we first identify the CADMM update in the matrix form as $\bm s^{k+1}=\bm D\bm s^k$ where $\bm D$ is fixed throughout the iterations. We then write the DQ-CADMM update as the sum of the ideal CADMM update plus an accumulated error term and finally utilize the linear convergence rate of the CADMM [cf. Equations (\ref{eqn:sx}) and (\ref{eqn:sa})]. In general, if the local objective functions do not have linear gradients or the linear convergence rate is not guaranteed (e.g., the LASSO is not differentiable and the corresponding CADMM update in this paper's fashion does not converge linearly), then the current proof no longer holds with deterministic quantization.
\end{remark}


\section{ADMM Based Algorithm for Quantized Consensus}
\label{sec:algorithm}
Let us summarize the two quantized versions of the CADMM: the PQ-CADMM converges linearly to the data average in the mean sense, but it does not guarantee a consensus within finite iterations; the DQ-CADMM, on the other hand, either converges to a consensus or cycles with the same mean of quantized variable values over one period at each node after a finite-time iteration, but results in an error from the average. 

As discussed in Remark \ref{rmk:notglobal}, we can first run the PQ-CADMM $2K$ times to obtain $\bar{x}_i=\frac{1}{K}\sum_{k=K+1}^{2K}x_{i}^k,$ which is a reasonable estimate of $x_\text{avg}$ at node $i$ according to Theorem \ref{thm:Qpresult}. Here $K$ can be chosen such that $\mathbb{E}[x^K_{i[Q_p]}]$ is close enough to $x_{\text{avg}}$ when we have the knowledge of agents' data and the network topology. Otherwise, we can simply set $K=\left\lceil10N\left(\log_{10}(\frac{1}{\Delta}+1)+1\right)\max\{-\log_{10}\rho,1\}\right\rceil$ or as large as permitted. Note also that $\bar{\alpha}_i = \frac{1}{K}\sum_{k=K+1}^{2K}\alpha_{i}^k$ is also a good estimate of $\alpha_i^*=r_i-x_{\text{avg}}$, and that $\bar{\bm\alpha}=[\bar{\alpha}_1;\bar{\alpha}_2;\cdots;\bar{\alpha}_N]=\frac{1}{K}\sum_{k=K+1}^{2K}\bm\alpha^k$ satisfies the initialization condition as $\bm\alpha^k$ lies in the column space of $\bm L_-$. We can therefore run the DQ-CADMM with this $\bar{x}_i$ and $\bar{\alpha}_i$ as initial values. The probabilistically quantized CADMM followed by deterministically quantized CADMM (PQDQ-CADMM) is presented in Algorithm \ref{tab:PQDQDCADMM}. 
\begin{algorithm}[htbp]
	\caption{PQDQ-CADMM for quantized consensus}
	\begin{algorithmic}[1]\label{tab:PQDQDCADMM}
	\REQUIRE Initialize~$\bm x^0=\bm 0$, $\bm\alpha^0=\bm 0$, and $\rho>0$. Set $K=\left\lceil10N\left(\log_{10}(\frac{1}{\Delta}+1)+1\right)\max\{-\log_{10}\rho,1\}\right\rceil$.
	\FOR{$k=0,1,\cdots,2K-1$, every node $i$}
	\STATE \begin{align} x_i^{k+1}&\gets\frac{1}{1+2\rho|\mathcal{N}_i|}\Bigg(\rho|\mathcal{N}_i|x_{i[Q_p]}^k+\rho\sum_{j\in\mathcal{N}_i} x_{j[Q_p]}^k\nn\\&~~~~-\alpha_i^k+r_i\Bigg),\nn\\
	\alpha_i^{k+1}&\gets\alpha_i^k+\rho\Bigg(|\mathcal{N}_i|x_{i[Q_p]}^{k+1}-\sum_{j\in\mathcal{N}_i} x_{j[Q_p]}^{k+1}\Bigg).\nn
	\end{align}
	\ENDFOR
	\STATE {\bf set} $\bm x^{2K}=\frac{1}{K}\sum_{l=K+1}^{2K} \bm x^l$, $\bm\alpha^{2K}=\frac{1}{K}\sum_{l=K+1}^{2K} \bm\alpha^l$, and $k=2K$.
	\REPEAT
			\STATE For $i=1,2,\cdots,N$, 
			\begin{align} x_i^{k+1}&\gets\frac{1}{1+2\rho|\mathcal{N}_i|}\Bigg(\rho|\mathcal{N}_i|x_{i[Q_d]}^k+\rho\sum_{j\in\mathcal{N}_i} x_{j[Q_d]}^k\nn\\&~~~~-\alpha_i^k+r_i\Bigg),\nn\\
			\alpha_i^{k+1}&\gets\alpha_i^k+\rho\Bigg(|\mathcal{N}_i|x_{i[Q_d]}^{k+1}-\sum_{j\in\mathcal{N}_i} x_{j[Q_d]}^{k+1}\Bigg).\nn
			\end{align}
			\STATE {\bf set} $k=k+1$.	
	\UNTIL{a predefined stopping criterion (e.g., a maximum iteration number) is satisfied.}
	\end{algorithmic}
\end{algorithm}
\section{Simulations}
\label{sec:simulation}
This section investigates the performance of the DQ-CADMM and the PQDQ-CADMM via numerical examples. Since existing methods with dithered quantization do not guarantee convergence to a consensus in finite iterations, we only compare our algorithms with those that uses deterministic quantization to reach a consensus, i.e., the gossip based method in \cite{Carli2010} and the classical method in \cite{Nedic2009}.

\subsection{Performance of the PQDQ-CADMM, the DQ-CADMM, the gossip based method, and the classical method}
\begin{figure}[htbp]
	\centering
	\includegraphics[width=\linewidth]{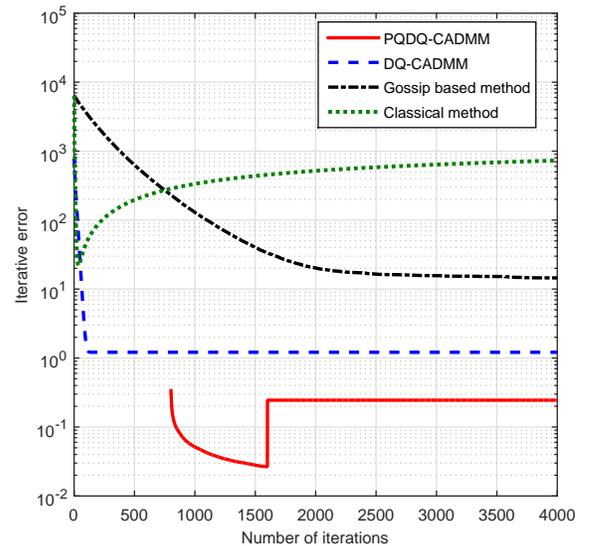}%
	\caption{Iterative error versus iterations where each plotted value is the average of $1000$ runs.}
	\label{fig:CEtra}
\end{figure}
  
 \begin{figure*}[ht]
	\centering
	\subfigure[]{%
		\includegraphics[width=0.325\linewidth]{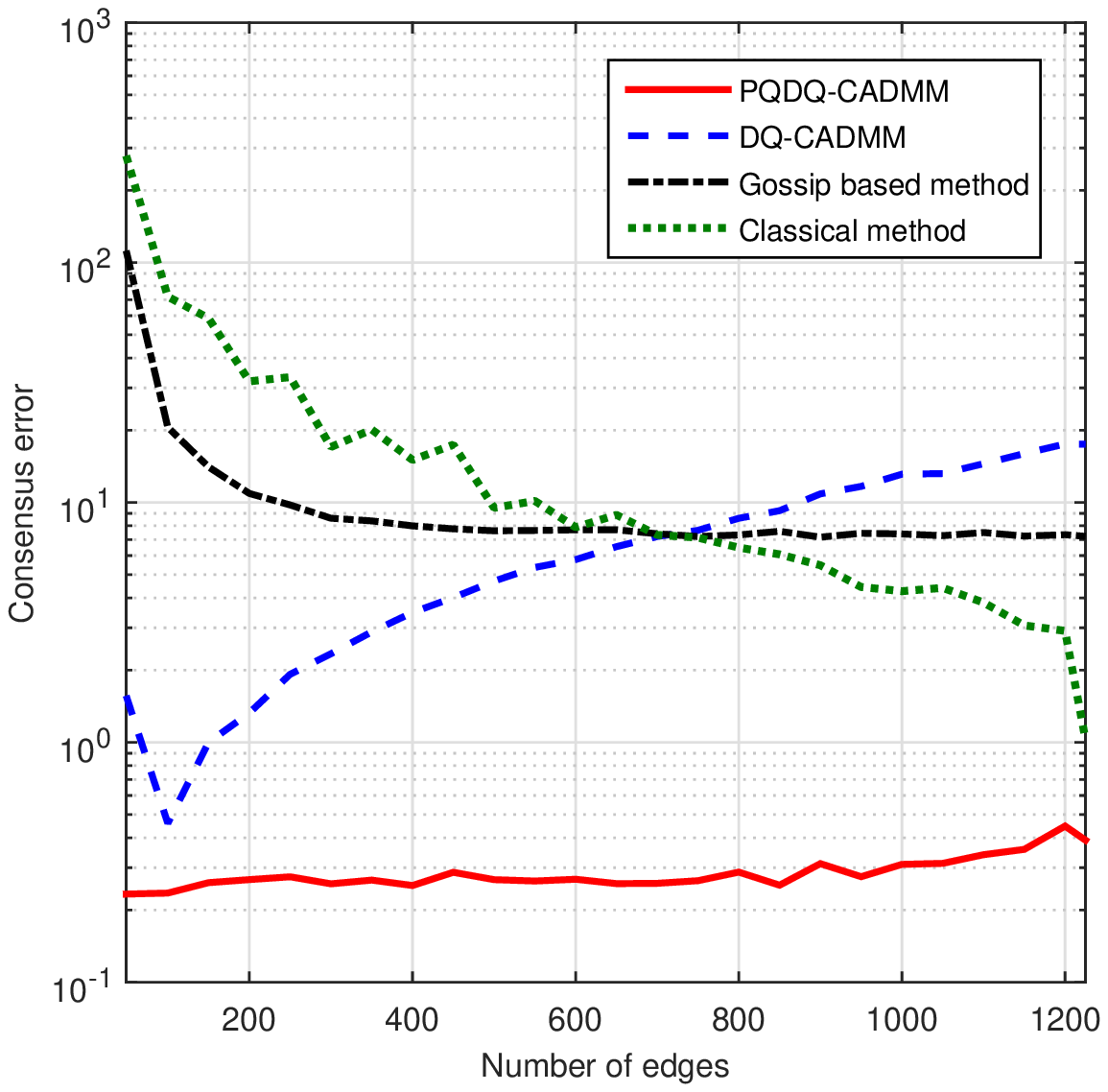}%
		\label{fig:cerra}%
	}
	\subfigure[]{%
		\includegraphics[width=0.325\linewidth]{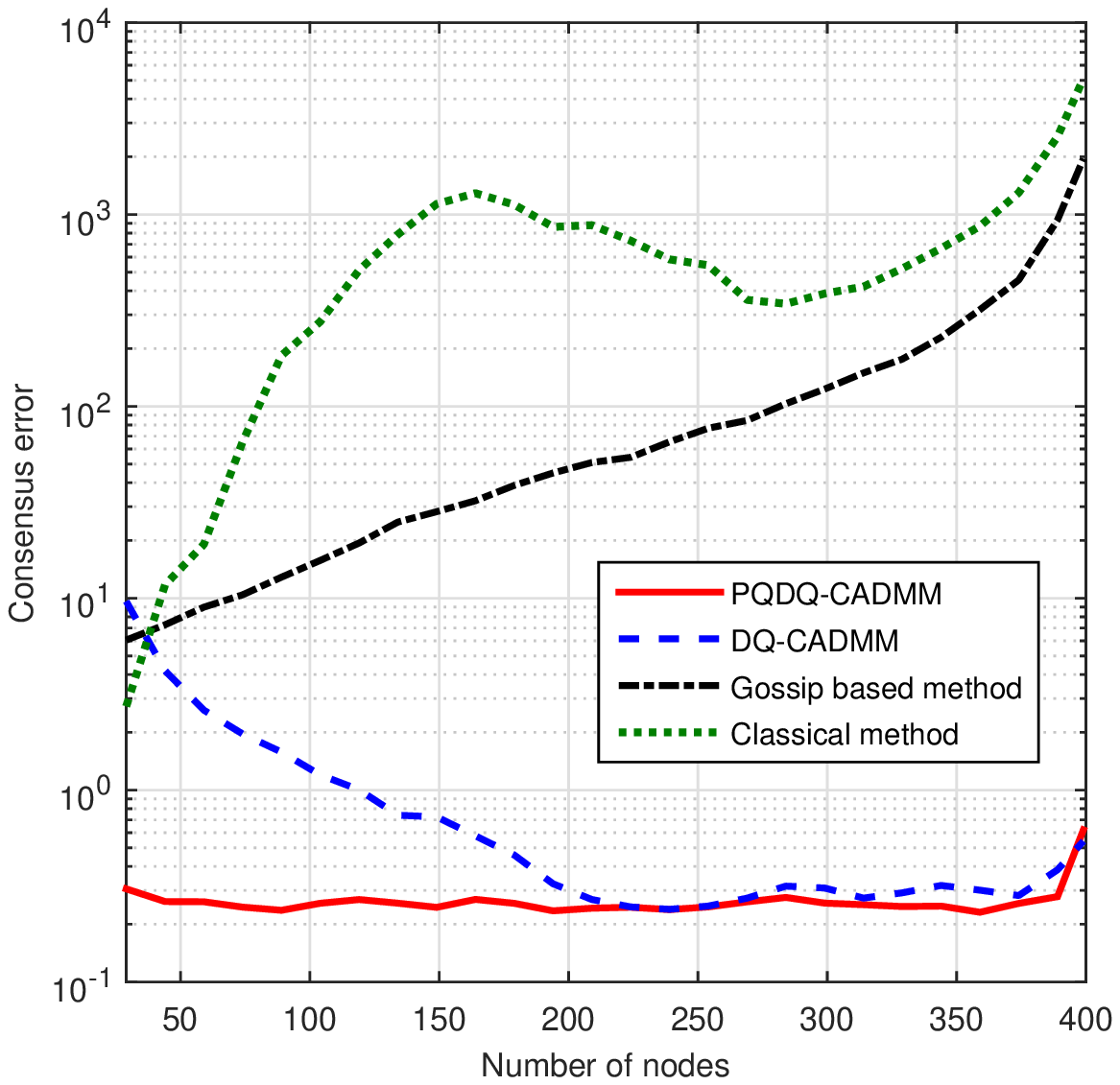}%
		\label{fig:cerrb}%
	}
	\subfigure[]{%
		\includegraphics[width=0.325\linewidth]{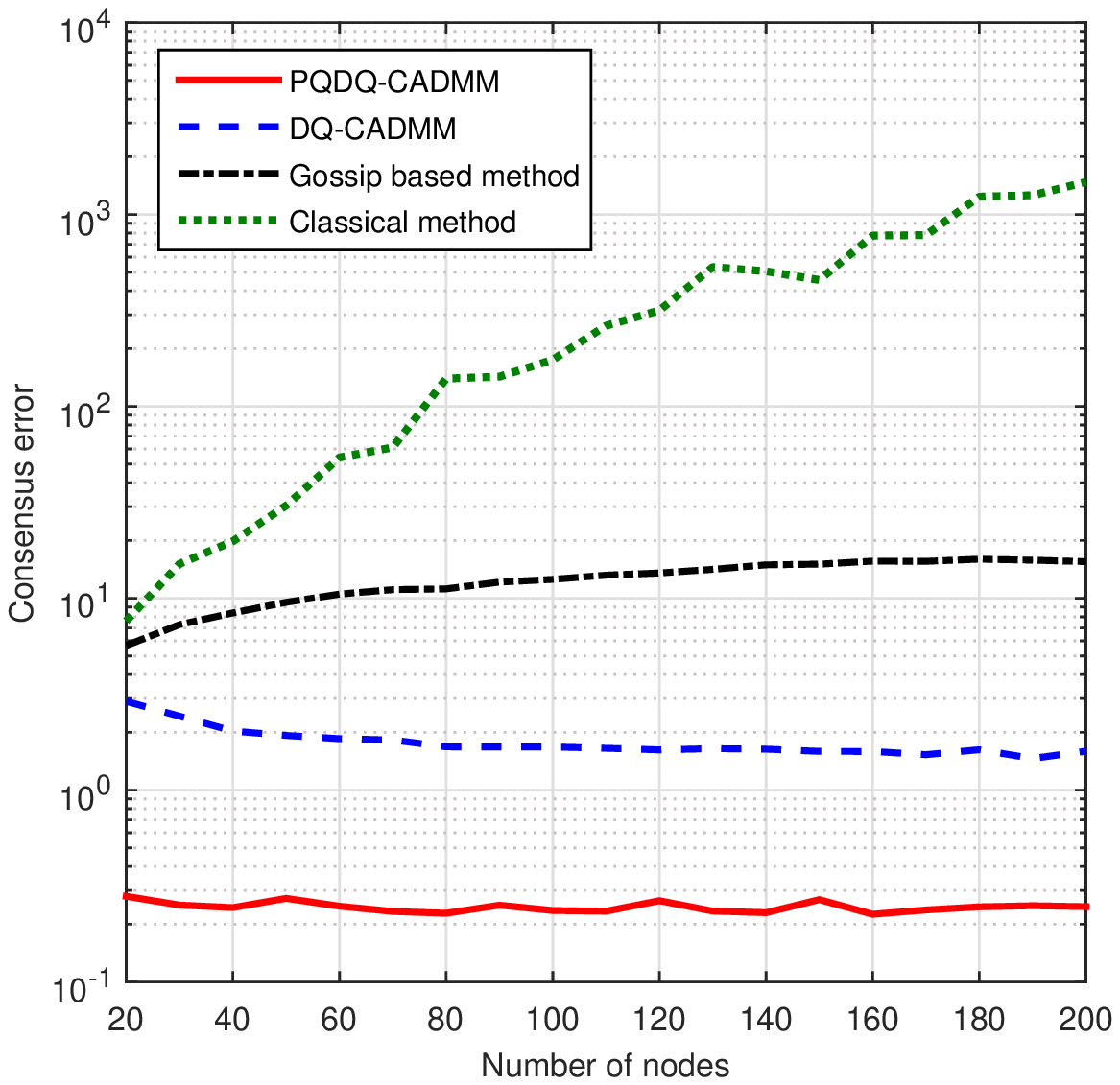}%
		\label{fig:cerrc}%
	}
	\caption{Consensus error of the four algorithms where $\Delta = 1$ and the plotted values are the average of $100$ runs; (a) fixing $N=50$ and varying $E\in[49, 1225]$, (b) fixing $E=400$ and varying $N\in[29, 399]$, (c) fixing $\frac{2E}{N}=10$ and varying $N\in[20,200]$.}
	\label{fig:cerr}
\end{figure*}
\begin{figure*}[ht]
	\centering
	\subfigure[]{%
		\includegraphics[width=.325\linewidth]{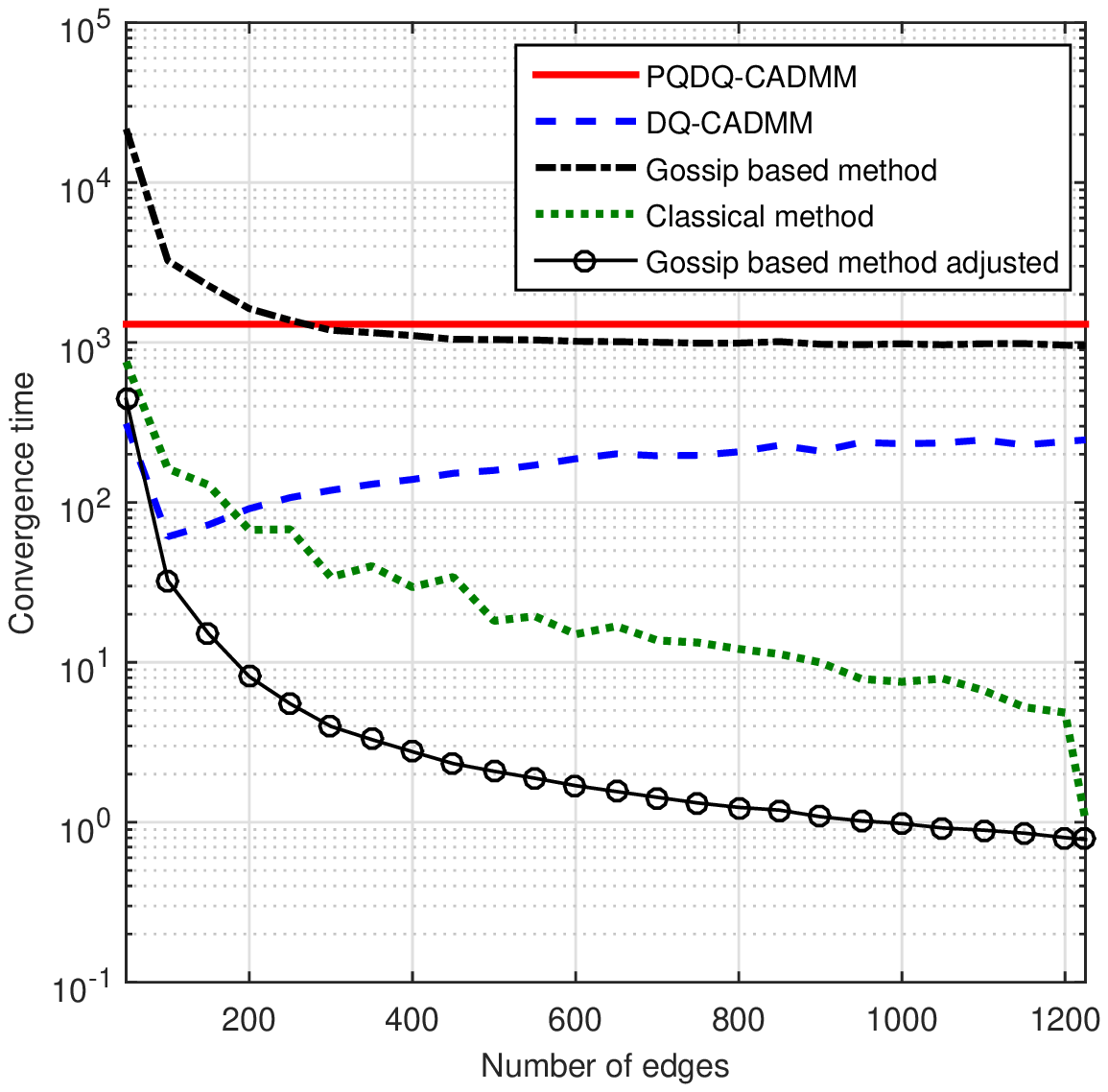}%
		\label{fig:ctimea}%
	}
		\subfigure[]{%
		\includegraphics[width=0.325\linewidth]{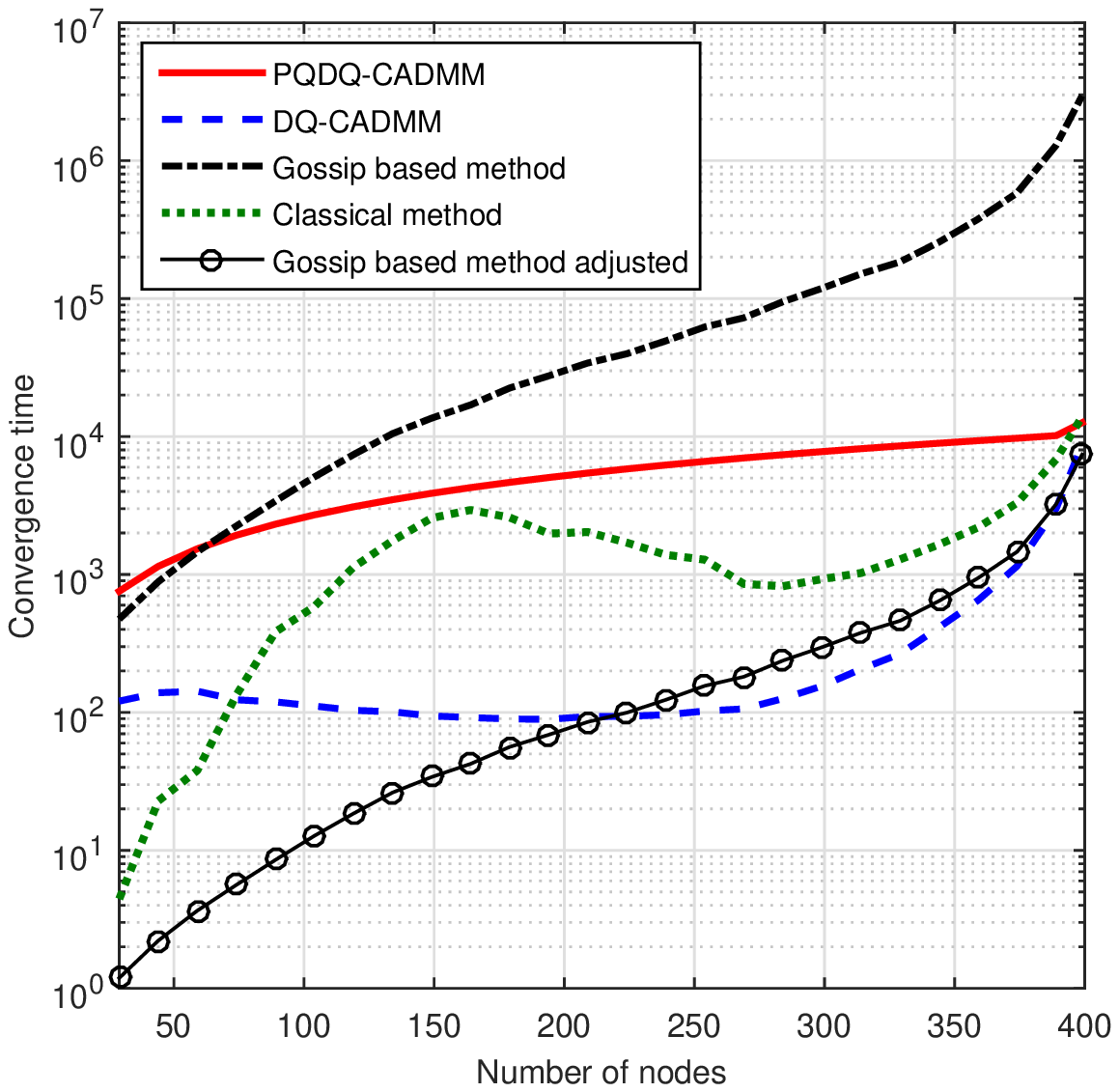}%
		\label{fig:ctimeb}%
	}
	\subfigure[]{%
		\includegraphics[width=.325\linewidth]{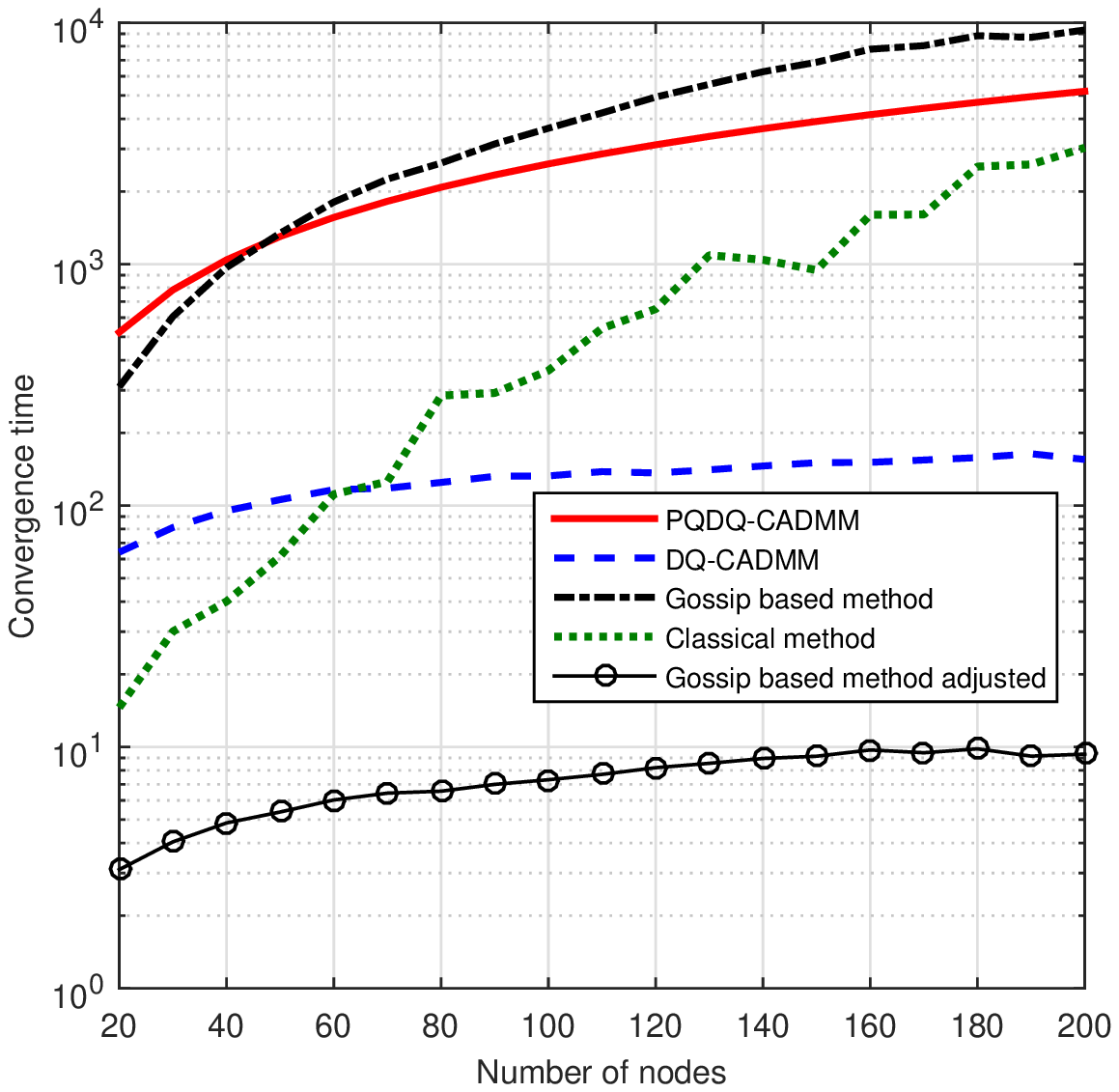}%
		\label{fig:ctimec}%
	}
	\caption{Convergence time of the four algorithms where $\Delta = 1$ and the plotted values are the average of $100$ runs; (a) $N=50$ and $E\in[49, 1225]$, (b) $E=400$ and $N\in[29, 399]$, (c) $\frac{2E}{N}=10$ and $N\in[20,200]$.}
	\label{fig:ctime}
\end{figure*}
To construct a connected graph with $N$ nodes and $E$ edges, we first generate a complete graph consisting of $N$ nodes, and then randomly remove $\frac{N(N-1)}{2}-E$ edges while ensuring that the network stays connected. Set $\Delta = 1$ and assume that agents' data have very high variances in large networks, e.g., let $r_i\sim\mathcal{N}(0,N^4)$. Our settings are
\begin{itemize}
\item PQDQ-CADMM: Set $\rho = 1$.
\item DQ-CADMM: Set $\rho = 1$, $\bm x^0 =\bm 0$ and $\bm\alpha^0=\bm 0$.
\item Gossip based method: We randomly pick one edge in $\mathcal{A}$ and perform the updating, i.e., if $(i,j)\in\mathcal{A}$ is chosen, then $x^{k+1}_i = x^{k+1}_j =\frac{1}{2}(x^{k}_{i[Q_d]}+x^k_{j[Q_d]})$. 
\item Classical method: Let $\bm W$ denote the weight matrix of the graph $\mathcal{G}_d=\{\mathcal{V},\mathcal{A}\}$. The updating rule is then given by $\bm x^{k+1} = \bm W \bm x^k_{[Q_{rd}]}$ where the subscript $_{[Q_{rd}]}$ denotes the rounding down quantization. We utilize the Metropolis weights defined in \cite{Xiao2005}: 
\begin{equation}
\begin{aligned}
W_{ij} =
\begin{cases}
(1+\max\{|\mathcal{N}_i|,|\mathcal{N}_j|\})^{-1}, &(i,j)\in\mathcal{A},\\
1-\sum_{k\in\mathcal{N}_i}W_{ik}, &i = j,\\
0, &\text{otherwise}.\nn
\end{cases}
\end{aligned}
\end{equation}
\end{itemize}

We simulate a connected network with $N=50$ nodes and $E=500$ edges. Define the iterative error as ${\|\bm x^k_{[Q]}-\bm 1x_{\text{avg}}\|_2}/{\sqrt{N}}$ which is equal to the consensus error $|x^*_{[Q]}-x_{\text{avg}}|$ when consensus is reached. Plotted in Fig.~\ref{fig:CEtra} is the iterative error of the four algorithms at every iteration $k$ with each value being the average of $1000$ runs. Note that we start the plot of the PQDQ-CADMM from the $(K+1)$th iteration as its first $K$ iterations are used only to reach a neighborhood of $x_{\text{avg}}$; at the $(2K+1)$th iteration, $\bm x^{2K+1}_{[Q]}$ is updated based on the running average of the $(K+1)$th iteration to the $2K$th iteration. The figure indicates that all the four algorithms converge to a consensus at one of the quantization levels. The average consensus error of the DQ-CADMM is $1.21$, which is much smaller than the upper bound $(\frac{1}{2}+\frac{2E}{N})\Delta = 20.5$. One can also see that the PQDQ-CADMM converges almost immediately after the $2K$th iteration. In the following we compare the consensus error and the convergence time of the four algorithms via simulations that respectively fix the number of nodes, the number of edges, and the average degree of the graph.

{\it Consensus error:}
In Fig.~\ref{fig:cerra} we fix $N=50$ and vary $E$ until the graph is complete. The gossip based method and the classical method have decreasing consensus errors as $E$ increases. The consensus error of the DQ-CADMM, however, becomes larger as the average degree and therefore the error bound increase. The PQDQ-CADMM has the smallest consensus error whose average of $100$ runs is less than $0.40$ for all $E$. We then fix $E = 400$ and let $N$ vary. Fig.~\ref{fig:cerrb} shows that the gossip based method and the classical method have increasing consensus errors as $N$ increases. The consensus error of the DQ-CADMM, on the contrary, decreases when $N$ becomes larger. The PQDQ-CADMM also has the smallest consensus error in this case. In the last setting we fix the average degree $\frac{2E}{N}=10$ while varying $N$. The classical method and the gossip based method then both have increasing consensus errors when $N$ and thus the range of agents' data increase. The consensus error of the DQ-CADMM is relatively small compared with the upper bound $(0.5+\frac{2E}{N})\Delta = 10.5$ and decreases when $N$ becomes larger. The proposed PQDQ-CADMM still has the smallest consensus error whose average of $100$ runs is less than $0.2$ for all $N$.  
 
We conclude that the consensus error of the gossip based method and the classical method depends on the average degree of the graph as well as the range of agents' data. Note that their consensus errors can be extremely large for a sparsely connected graph. The DQ-CADMM has an increasing consensus error when the average degree increases while the PQDQ-CADMM performs almost the same for all network structures in terms of the consensus error. 

{\it Convergence time:} We study the convergence time of the four algorithms via numerical examples in Fig.~\ref{fig:ctime}. Since the gossip based method involves only one edge and the other three methods utilize all the edges at each iteration, we plot also the quotient of the convergence time of the gossip based method divided by the number of edges, namely, Gossip based method adjusted, in the figure. 

In Fig.~\ref{fig:ctimea}, the gossip based method and the classical method converge slower as the graph becomes sparser. When the average degree is fixed, they have longer convergence time as $N$ increases. Therefore, the convergence time of the gossip based method and the classical method is also affected by the average degree of the graph and the range of agents' data. Different from the gossip based and classical methods, we see in Fig.~\ref{fig:ctimea} that the convergence time of the DQ-CADMM increase as the graph becomes denser. In Fig.~\ref{fig:ctimeb} and Fig.~\ref{fig:ctimec}, however, the convergence time also increases while the graph becomes sparser, which is possibly because of the increased distance between starting points and optimal values. Exactly characterizing the convergence time of the DQ-CADMM is beyond the scope of the current paper and will be treated as future work. For the PQDQ-CADMM, we observe that the significant portion of its convergence time is spent on achieving an approximate estimate of $x_\text{avg}$, i.e., running the PQ-CADMM with $2K$ iterations. With good starting points, the DQ-CADMM converges almost immediately. 

\subsection{Performance of the PQDQ-CADMM with different quantization resolutions}
\begin{figure}[h]
\vspace{-0.15in}
	\centering
	\includegraphics[width=\linewidth]{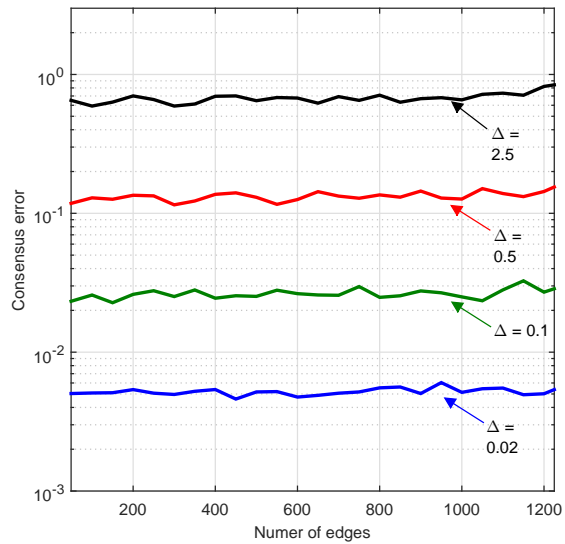}%
	\caption{Consensus error of the PQDQ-CADMM with different quantization resolutions, i.e., $\Delta\in\{0.02, 0.1, 0.5, 2.5\}$, for $N=50$ and $E\in[49,1225]$; each plotted value is the average of $100$ runs.}
	\label{fig:CEDelta}
\end{figure}

We next consider the effect of the quantization resolution on the PQDQ-CADMM. Fig.~\ref{fig:CEDelta} plots consensus errors of the PQDQ-CADMM with $N = 50$ and $E\in[49,1225]$ for $\Delta\in\{0.02, 0.1, 0.5, 2.5\}$. The consensus error tends to increase on the average as the quantization resolution becomes larger, which is not surprising since a coarse quantization indicates a higher loss of information at each update. We then calculate the ratio of the consensus error to the quantization resolution: the plotted values, which are the averages of $100$ runs, all lie in $(0.227\Delta, 0.337\Delta)$ and the variances are less than $0.051$. Moreover, the convergence time of each quantization resolution has a mean of $(2K+2.1)$ iterations and a variance less than $0.0008$, which coincides with our previous analysis that the PQDQ-CADMM converges immediately after the first $2K$ iterations.

\subsection{Cyclic Cases}
\begin{figure}[h]
    \centering
    {\includegraphics[width=1.0\linewidth,height=0.85\linewidth]{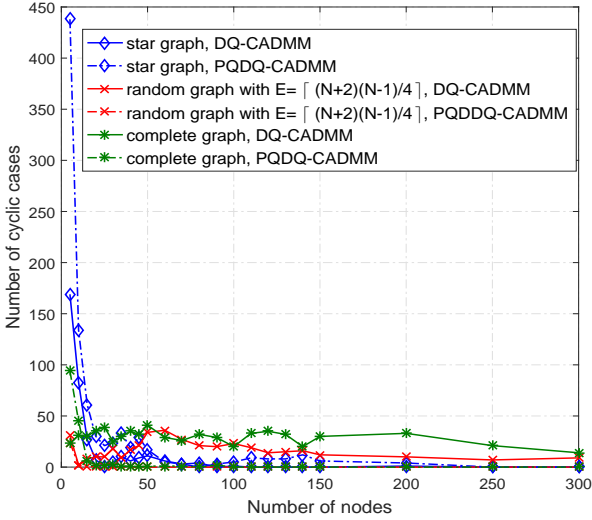}}
      \caption{Number of cyclic cases in $10^4$ trials.}
      \label{fig:cycnum}
\end{figure}

While we prove that the DQ-CADMM either converges or cycles in Theorem~4, it is noted that the above numerical examples all lead to reach convergence results. Indeed, the proposed deterministic algorithms, the DQ-CADMM and PQDQ-CADMM, converges in most cases as shown by the following simulation. For connected networks with $N$ nodes, we consider star graph which has the smallest average degree, randomly generated graph that has intermediate average degree, and complete graph that has the largest average degree. The result is given in Fig.~\ref{fig:cycnum} where the $y$-axis represents the number of cyclic cases in $10^4$ trials. Clearly, the DQ-CADMM and PQDQ-CADMM with fixed parameter $\rho=1$ converge in most cases, particularly with large networks. 
\vspace{-0.15in}

\section{Conclusion}
\label{sec:conclusion}
In this paper we have proposed an efficient algorithm, the PQDQ-CADMM, for quantized consensus problems. We first study the effects of both probabilistic and deterministic quantizations on the CADMM. With probabilistic quantization, the PQ-CADMM converges linearly to the data average in the mean sense. In the deterministic case, we can bound the sum of the absolute value of each error term caused by quantization using the global and linear convergence of the CADMM and thus prove that the DQ-CADMM either converges or cycles. We finally combine the two quantized versions of the CADMM to obtain the PQDQ-CADMM algorithm, where the PQ-CADMM to is used to get an initial estimate of the data average and the DQ-CADMM is used subsequently for consensus reaching purpose. Simulations show that our PQDQ-CADMM provides the best result than all existing methods using deterministic quantization in terms of the consensus error.

Our approach also motivates a number of further research directions:
\begin{enumerate}
\item Data communications between agents were assumed to be perfect in this paper. In practice, channel impairment may lead to imperfect transmissions. Moreover, the links between agents may fail and the network topology may vary randomly, as studied in \cite{Nedic2009,Zhu2009,Kar2010}. It is thus meaningful to investigate how our algorithm performs in these settings.
\item The algorithm parameter $\rho$ is another interesting topic in the DQ-CADMM. Roughly speaking, a smaller $\rho$ may result in a small consensus error but a longer time to reach the convergent or cyclic result. Therefore, tts choice should be guided depending on whether a small consensus error or fast consensus speed is desired.
\item We only considered the unbounded quantization scheme in this paper. It is also interesting to consider bounded quantization that is used in many applications as it significantly reduces the amount of data that needs to be exchanged.
\end{enumerate}
\appendices
\section*{Appendix}
\begin{IEEEproof}[Proof of Lemma \ref{lem:linearconvergence}] We first manipulate (\ref{eqn:admmupdates}) to derive equivalent updates 
\begin{align}
\label{eqn:admmupdatesequivalent1}
&\nabla f(\bm x^{k+1}) + \bm A^T\bm\lambda^k+\rho \bm A^T\bm B(\bm z^k-\bm z^{k+1})=\bm 0,\\
\label{eqn:admmupdatesequivalent2}
&\bm B^T\bm\lambda^{k+1} = \bm 0,\\
\label{eqn:admmupdatesequivalent3}
&\bm\lambda^{k+1}-\bm\lambda^k-\rho(\bm A \bm x^{k+1}+\bm B \bm z^{k+1})=\bm 0,
\end{align}
where (\ref{eqn:admmupdatesequivalent1}) and (\ref{eqn:admmupdatesequivalent2}) are from multiplying the two sides of the $\bm \lambda$-update by $\bm A^T$ and $\bm B^T$ and adding them to the $\bm x$-update and $\bm z$-update, respectively. Recalling $\bm \lambda=[\bm\beta;\bm\gamma]$ with $\bm\beta,\bm \gamma\in\mathbb{R}^{2E}$ and $\bm B = [-\bm I_{2E};-\bm I_{2E}]$, we know that $\bm\beta^{k+1}=-\bm\gamma^{k+1}$ from (\ref{eqn:admmupdatesequivalent2}). Since we initialize $\bm\beta^0=-\bm\gamma^0$, we have $\bm B^T\bm\lambda^k=\bm 0$ for $k = 0,1,\cdots$. Equation (\ref{eqn:admmupdatesequivalent1}) then reduces to $\nabla f(\bm x^{k+1})+\bm M_-\bm\beta^{k+1}-\rho\bm M_+(\bm z^k-\bm z^{k+1})=\bm 0$, and (\ref{eqn:admmupdatesequivalent3}) splits into $\bm\beta^{k+1}-\bm\beta^k-\rho \bm A_1 \bm x^{k+1}+\rho \bm z^{k+1}=\bm 0$ and $\bm\gamma^{k+1}-\bm\gamma^k-\rho\bm A_2\bm x^{k+1}+\rho\bm z^{k+1}=\bm 0$. Summing and subtracting these two equations we have $\frac{1}{2}\bm M_+^T\bm x^{k+1}-\bm z^{k+1}=\bm 0$ and $\bm\beta^{k+1}-\bm\beta^k-\frac{\rho}{2}\bm M_-^T\bm x^{k+1}=\bm 0$. With the initialization $\bm z^0=\frac{1}{2}\bm M_+^T\bm x^0$, $\frac{1}{2}\bm M_+^T\bm x^{k}-\bm z^{k}=\bm 0$ holds true for $k=0,1,\cdots$. Since $\bm x^*$ is unique and equal to $\bm 1x_\text{avg}$ according to Lemma \ref{lem:globalconvergence}, $\bm z^*=\frac{1}{2}\bm M_+^T\bm x^*$ is also unique. To summarize, with the initialization $\bm\beta^0=-\bm\gamma^0$ and $\bm z^0=\frac{1}{2}\bm M_+^T\bm x^0$, (\ref{eqn:admmupdatesequivalent1})-(\ref{eqn:admmupdatesequivalent3}) reduce to 
\begin{align}
\label{eqn:admmupdates21}
&\nabla f(\bm x^{k+1})+\bm M_-\bm\beta^{k+1}-\rho \bm M_+(\bm z^k-\bm z^{k+1})=\bm 0,\\
\label{eqn:admmupdates22}
&\bm\beta^{k+1}-\bm\beta^k-\frac{\rho}{2}\bm M_-^T\bm x^{k+1}=\bm 0,\\
\label{eqn:admmupdates23}
&\frac{1}{2}\bm M_+^T\bm x^{k+1}-\bm z^{k+1}=\bm 0,
\end{align}
which further lead to $\bm x^k\to \bm x^*=\bm 1x_\text{avg}$ and $\bm z^k\to \frac{1}{2}\bm M_+^T\bm x^*= \frac{1}{2}\bm M_+^T\bm 1 x_\text{avg}$ uniquely as $k\to\infty$.
Taking $k\to\infty$ in (\ref{eqn:admmupdates21})-(\ref{eqn:admmupdates23}) and using global convergence, we get
\begin{align}
\label{eqn:admmupdates31}
&\nabla f(\bm x^*)+\bm M_-\bm\beta^*=\bm 0,\\
\label{eqn:admmupdates32}
&\bm M_-^T\bm x^{*}=\bm 0,\\
\label{eqn:admmupdates33}
&\frac{1}{2}\bm M_+^T\bm x^*-\bm z^*=\bm 0.
\end{align}
We can now use (\ref{eqn:admmupdates31}) to demonstrate the uniqueness of $\bm\beta^*$ if we also initialize $\bm\beta_0$ in the column space of $\bm M_-^T$. Note that if $\bm\beta^0$ lies in the column space of $\bm M_-^T$ then (\ref{eqn:admmupdates22}) indicates that $\bm\beta^k$ also lies in the column space of $\bm M_-^T$, $k=0,1,\cdots$. The uniqueness of $\bm\beta^*$ then follows from the uniqueness of $\bm x^*$ and Lemma \ref{lem:abrelation}.

Next we show the linear convergence of $\bm u^k$. Subtracting (\ref{eqn:admmupdates21})-(\ref{eqn:admmupdates23}) from (\ref{eqn:admmupdates31})-(\ref{eqn:admmupdates33}), respectively, and using $\nabla f(\bm x)=\bm x-\bm r$, we have
\begin{align}
\label{eqn:admmupdatesfinal}
&\bm x^{k+1}-\bm x^*=\rho\bm M_+^T(\bm z^k-\bm z^{k+1})-\bm M_-(\bm \beta^{k+1}-\bm\beta^*),\\
\label{eqn:admmupdatesfinal1}
&\frac{\rho}{2}\bm M_-^T(\bm x^{k+1}-\bm x^*)=\bm\beta^{k+1}-\bm\beta^k,\\
\label{eqn:admmupdatesfinal2}
&\frac{1}{2}\bm M_+^T(\bm x^{k+1}-\bm x^*)=\bm z^{k+1}-\bm z^*.
\end{align}
We therefore obtain 
\begin{align}
&\hspace{-0.08in}~~~~\|\bm x^{k+1}-\bm x^*\|_2^2\nn\\
&\hspace{-0.08in}\stackrel{(a)}{=}\langle \bm x^{k+1}-\bm x^*,\rho\bm M_+^T(\bm z^k-\bm z^{k+1})-\bm M_-(\bm\beta^{k+1}-\bm\beta^*)\rangle\nn\\
&\hspace{-0.08in}\stackrel{(b)}{=}2\rho\langle \bm z^k-\bm z^{k+1},\bm z^{k+1}-\bm z^* \rangle + \frac{2}{\rho} \langle\bm \beta^k-\bm\beta^{k+1},\bm\beta^{k+1}-\bm\beta^*\rangle\nn\\
\label{eqn:x13}
&\hspace{-0.08in}\stackrel{(c)}{=}\|\bm u^k-\bm u^*\|_{\bm G}^2-\|\bm u^{k+1}-\bm u^*\|_{\bm G}^2-\|\bm u^k-\bm u^{k+1}\|_{\bm G}^2,
\end{align}
where $(a)$ is from (\ref{eqn:admmupdatesfinal}), $(b)$ is from (\ref{eqn:admmupdatesfinal1}) and (\ref{eqn:admmupdatesfinal2}), and $(c)$ is from the definitions of $\bm u$ and $\bm G$. Due to (\ref{eqn:x13}), to prove (\ref{eqn:Ulinearconvergence}) we only need to show
\begin{equation}
\begin{aligned}
\|\bm u^k-\bm u^{k+1}\|_{\bm G}^2 + \|\bm x^{k+1}-\bm x^*\|_2^2\geq\delta\|\bm u^{k+1}-\bm u^*\|_{\bm G}^2,\nn
\end{aligned}
\end{equation}
which is equivalent to 
\begin{equation}
\begin{aligned}
\rho\|\bm z^{k+1}-\bm z^k\|_2^2+\frac{1}{\rho}&\|\bm\beta^{k+1}-\bm\beta^k\|_2^2+\|\bm x^{k+1}-\bm x^*\|_2^2\nn\\&\geq\delta\rho\|\bm z^{k+1}-\bm z^*\|_2^2+\frac{\delta}{\rho}\|\bm\beta^{k+1}-\bm\beta^*\|_2^2.\nn
\end{aligned}
\end{equation}
It then suffices to show
\begin{align}
\rho\|\bm z^{k+1}-&\bm z^k\|_2^2+\|\bm x^{k+1}-\bm x^*\|_2^2\nn\\&
\label{eqn:x41}
\geq\delta\rho\|\bm z^{k+1}-\bm z^*\|_2^2+\frac{\delta}{\rho}\|\bm\beta^{k+1}-\bm\beta^*\|_2^2.
\end{align}
The rest of this proof is to establish that $\delta\rho\|\bm z^{k+1}-\bm z^*\|_2^2$ and $\frac{\delta}{\rho}\|\bm\beta^{k+1}-\bm\beta^*\|_2^2$ are upper bounded by two non-overlapping parts of the left side of (\ref{eqn:x41}), respectively.  

We first have from (\ref{eqn:admmupdatesfinal2}) that
\begin{align}
\label{eqn:x5}
\|\bm z^{k+1}-\bm z^*\|_2^2&=\frac{1}{4}\|\bm M_+^T(\bm x^{k+1}-\bm x^*)\|^2_2\nn\\&\leq\frac{1}{4}\sigma_{\max}^2(\bm M_+)\|\bm x^{k+1}-\bm x^*\|_2^2.
\end{align}
To upper bound $\|\bm\beta^{k+1}-\bm\beta^*\|_2^2$, we first notice that $\bm\beta^{k+1}-\bm\beta^*$ lies in the column space of $\bm M_-^T$. Therefore, 
\begin{align}
\label{eqn:beatalphal}
\|\bm M_-(\bm\beta^{k+1}-\bm\beta^*)\|_2^2\geq{\tilde{\sigma}}_{\min}^2(\bm M_-)\|\bm\beta^{k+1}-\bm\beta^*\|_2^2.
\end{align}
Now using (\ref{eqn:beatalphal}) and (\ref{eqn:admmupdatesfinal}) we get 
\begin{align}
&~~~~\|\bm \beta^{k+1}-\bm \beta^*\|_2^2\nn\\
&\leq\frac{1}{{\tilde{\sigma}}_{\min}^2(\bm M_-)}\|\bm M_-(\bm\beta^{k+1}-\bm\beta^*)\|_2^2\nn\\
&=\frac{1}{{\tilde{\sigma}}_{\min}^2(\bm M_-)}\|(\bm x^{k+1}-\bm x^*)-\rho \bm M_+(\bm z^k-\bm z^{k+1})\|_2^2\nn\\
&\stackrel{(a)}{\leq} \frac{2}{{\tilde{\sigma}}_{\min}^2(\bm M_-)}\left(\|\bm x^{k+1}-\bm x^*\|_2^2+\rho^2\|\bm M_+(\bm z^k-\bm z^{k+1})\|_2^2\right)\nn\\
\label{eqn:x62}
&\leq\frac{2}{{\tilde{\sigma}}_{\min}^2(\bm M_-)}\big(\|\bm x^{k+1}-\bm x^*\|_2^2\nn\\
&\hspace{1.2in}+\rho^2\sigma_{\max}^2(\bm M_+)\|\bm z^{k+1}-\bm z^k\|_2^2\big),
\end{align}
where $(a)$ is from the Cauchy-Schwarz inequality together with the fact $2p_1p_2\leq p_1^2+p_2^2$ for any $p_1,p_2\in\mathbb{R}$.
Combining (\ref{eqn:x5}) and (\ref{eqn:x62}), we have 
\begin{align}
&~\rho\frac{2\sigma_{\max}^2(\bm M_+)}{{\tilde{\sigma}}_{\min}^2(\bm M_-)}\|\bm z^{k+1}-\bm z^k\|_2^2\nn\\&+\left(\frac{1}{4}\rho\sigma_{\max}^2(\bm M_+)+\frac{2}{{\rho\tilde{\sigma}}_{\min}^2(\bm M_-)}\right)\|\bm x^{k+1}-\bm x^*\|_2^2\nn\\
&\hspace{1.3in}\geq\rho\|\bm z^{k+1}-\bm z^*\|_2^2+\frac{1}{\rho}\|\bm\beta^{k+1}-\bm\beta^*\|_2^2.\nn
\end{align}
The proof is thus complete by picking 
$$\delta = \min\left\{\frac{{\tilde{\sigma}}_{\min}^2(\bm M_-)}{2\sigma_{\max}^2(\bm M_+)}, \frac{4\rho{\tilde{\sigma}}_{\min}^2(\bm M_-)}{\rho^2\sigma_{\max}^2(\bm M_+){\tilde{\sigma}}_{\min}^2(\bm M_-)+ 8}\right\}.$$
\end{IEEEproof}


\section*{Acknowledgments}
The authors would like to thank Professor Zhi-Quan Luo, from University of Minnesota, Professor Mingyi Hong, from Iowa State University, and Professor Lixin Shen, from Syracuse University, for helpful discussions.

\bibliographystyle{IEEEtran}
\bibliography{SZhuBib}

\begin{thebibliography}{10}
\providecommand{\url}[1]{#1}
\csname url@samestyle\endcsname
\providecommand{\newblock}{\relax}
\providecommand{\bibinfo}[2]{#2}
\providecommand{\BIBentrySTDinterwordspacing}{\spaceskip=0pt\relax}
\providecommand{\BIBentryALTinterwordstretchfactor}{4}
\providecommand{\BIBentryALTinterwordspacing}{\spaceskip=\fontdimen2\font plus
\BIBentryALTinterwordstretchfactor\fontdimen3\font minus
  \fontdimen4\font\relax}
\providecommand{\BIBforeignlanguage}[2]{{%
\expandafter\ifx\csname l@#1\endcsname\relax
\typeout{** WARNING: IEEEtran.bst: No hyphenation pattern has been}%
\typeout{** loaded for the language `#1'. Using the pattern for}%
\typeout{** the default language instead.}%
\else
\language=\csname l@#1\endcsname
\fi
#2}}
\providecommand{\BIBdecl}{\relax}
\BIBdecl

\bibitem{Zhu2015a}
S.~Zhu and B.~Chen, ``Distributed average consensus with deterministic
  quantization: {an ADMM} approach,'' in \emph{Proc. IEEE Global Conf. Signal
  and Information Processing (GlobalSIP)}, Orlando, FL, Dec. 2015.

\bibitem{Ren2007}
W.~Ren, R.~W. Beard, and E.~M. Atkins, ``Information consensus in multivehicle
  cooperative control,'' \emph{IEEE Control Systems}, vol.~27, no.~2, pp.
  71--82, Apr. 2007.

\bibitem{Cao2013}
Y.~Cao, W.~Yu, W.~Ren, and G.~Chen, ``An overview of recent progress in the
  study of distributed multi-agent coordination,'' \emph{IEEE Trans. Ind.
  Inf.}, vol.~9, no.~1, pp. 427--438, Feb. 2013.

\bibitem{Lynch1996distributed}
N.~A. Lynch, \emph{Distributed Algorithms}.\hskip 1em plus 0.5em minus
  0.4em\relax San Francisco, CA: Morgan Kaufmann, 1996.

\bibitem{Ren2005}
W.~Ren and R.~W. Beard, ``Consensus seeking in multiagent systems under
  dynamically changing interaction topologies,'' \emph{IEEE Trans. Autom.
  Control}, vol.~50, no.~5, pp. 655--661, May 2005.

\bibitem{Xiao2005}
L.~Xiao, S.~Boyd, and S.~Lall, ``A scheme for robust distributed sensor fusion
  based on average consensus,'' in \emph{Proc. Int. Symp. Information
  Processing in Sensor Networks}, Los Angeles, CA, Apr. 2005.

\bibitem{Xu1996}
C.~Xu and F.~C. Lau, \emph{Load Balancing in Parallel Computers: Theory and
  Practice}.\hskip 1em plus 0.5em minus 0.4em\relax Dordrecht, Germany: Kluwer,
  1997.

\bibitem{Kashyap2007}
A.~Kashyap, T.~Ba{\c{s}}ar, and R.~Srikant, ``Quantized consensus,''
  \emph{Automatica}, vol.~43, no.~7, pp. 1192--1203, 2007.

\bibitem{Xiao2004}
L.~Xiao and S.~Boyd, ``Fast linear iterations for distributed averaging,''
  \emph{Syst. Contr. Lett.}, vol.~53, no.~1, pp. 65--78, 2004.

\bibitem{Jakovetic2010}
D.~Jakoveti{\'{c}}, J.~Xavier, and J.~M.~F. Moura, ``Weight optimization for
  consensus algorithms with correlated switching topology,'' \emph{IEEE Trans.
  Signal Process.}, vol.~58, no.~7, pp. 3788--3801, Jul. 2010.

\bibitem{Nedic2009}
A.~Nedic, A.~Olshevsky, A.~Ozdaglar, and J.~N. Tsitsiklis, ``On distributed
  averaging algorithms and quantization effects,'' \emph{IEEE Trans. Autom.
  Control}, vol.~54, no.~11, pp. 2506--2517, Nov. 2009.

\bibitem{Aysal2009}
T.~C. Aysal, M.~E. Yildiz, A.~D. Sarwate, and A.~Scaglione, ``Broadcast gossip
  algorithms for consensus,'' \emph{IEEE Trans. Signal Process.}, vol.~57,
  no.~7, pp. 2748--2761, Jul. 2009.

\bibitem{Boyd2006}
S.~Boyd, A.~Ghosh, B.~Prabhakar, and D.~Shah, ``Randomized gossip algorithms,''
  \emph{IEEE Trans. Inf. Theory}, vol.~52, no.~6, pp. 2508--2530, Jun. 2006.

\bibitem{Schizas2008}
I.~D. Schizas, A.~Ribeiro, and G.~B. Giannakis, ``Consensus in {\it{ad hoc}}
  {WSN}s with noisy links---{Part I}: Distributed estimation of deterministic
  signals,'' \emph{IEEE Trans. Signal Process.}, vol.~56, no.~1, pp. 350--364,
  Jan. 2008.

\bibitem{Zhu2009}
H.~Zhu, G.~B. Giannakis, and A.~Cano, ``Distributed in-network channel
  decoding,'' \emph{IEEE Trans Signal Process}, vol.~57, no.~10, pp.
  3970--3983, Oct. 2009.

\bibitem{Erseghe2011}
T.~Erseghe, D.~Zennaro, E.~Dall'Anese, and L.~Vangelista, ``Fast consensus by
  the alternating direction multipliers method,'' \emph{IEEE Trans. Signal
  Process.}, vol.~59, no.~11, pp. 5523--5537, Nov. 2011.

\bibitem{Aysal2008}
T.~C. Aysal, M.~J. Coates, and M.~G. Rabbat, ``Distributed average consensus
  with dithered quantization,'' \emph{IEEE Trans. Signal Process.}, vol.~56,
  no.~10, pp. 4905--4918, Oct. 2008.

\bibitem{Kar2010}
S.~Kar and J.~M. Moura, ``Distributed consensus algorithms in sensor networks:
  Quantized data and random link failures,'' \emph{IEEE Trans. Signal
  Process.}, vol.~58, no.~3, pp. 1383--1400, 2010.

\bibitem{Chamie2014}
M.~E. Chamie, J.~Liu, and T.~Ba{\c{s}}ar, ``Design and analysis of distributed
  averaging with quantized communication,'' in \emph{Proc. 53rd IEEE Conf.
  Decision and Control}, Los Angeles, CA, Dec. 2014.

\bibitem{Carli2010}
R.~Carli, F.~Fagnani, P.~Frasca, and S.~Zampieri, ``Gossip consensus algorithms
  via quantized communication,'' \emph{Automatica}, vol.~46, no.~1, pp. 70--80,
  2010.

\bibitem{Tsitsiklis1984}
J.~N. Tsitsiklis, ``Problems in decentralized decision making and
  computation,'' DTIC Document, Tech. Rep., 1984.

\bibitem{BoydADMM}
S.~Boyd, N.~Parikh, E.~Chu, B.~Peleato, and J.~Eckstein, ``Distributed
  optimization and statistical learning via the alternating direction method of
  multipliers,'' \emph{Foundations and Trends in Machine Learning}, vol.~3,
  no.~1, pp. 1--122, 2011.

\bibitem{Schuchman1964}
L.~Schuchman, ``Dither signals and their effect on quantization noise,''
  \emph{IEEE T. Commun. Techn.}, vol.~12, no.~4, pp. 162--165, Dec. 1964.

\bibitem{He2012}
B.~He and X.~Yuan, ``On the {$O(1/n)$} convergence rate of the douglas-rachford
  alternating direction method,'' \emph{SIAM J. Numer. Anal.}, vol.~50, no.~2,
  pp. 700--709, 2012.

\bibitem{Hong2012}
M.~Hong and Z.-Q. Luo, ``On the linear convergence of the alternating direction
  method of multipliers,'' \emph{arXiv preprint arXiv:1208.3922}, 2012.

\bibitem{Deng2016}
W.~Deng and W.~Yin, ``On the global and linear convergence of the generalized
  alternating direction method of multipliers,'' \emph{J. Sci. Comput.},
  vol.~66, no.~3, pp. 889--916, 2016.

\bibitem{Xiao2005a}
J.-J. Xiao and Z.-Q. Luo, ``Decentralized estimation in an inhomogeneous
  sensing environment,'' \emph{IEEE Trans. Inf. Theory}, vol.~51, no.~10, pp.
  3564--3575, Oct. 2005.

\bibitem{Shi2014}
W.~Shi, Q.~Ling, K.~Yuan, G.~Wu, and W.~Yin, ``On the linear convergence of the
  admm in decentralized consensus optimization,'' \emph{IEEE Trans. Signal
  Process.}, vol.~62, no.~7, pp. 1750--1761, Apr. 2014.

\bibitem{ChungSpectral}
F.~R. Chung, \emph{Spectral Graph Theory}.\hskip 1em plus 0.5em minus
  0.4em\relax American Mathematical Soc., 1997, vol.~92.

\bibitem{Ghadimi2015ParaSel}
E.~Ghadimi, A.~Teixeira, I.~Shames, and M.~Johansson, ``Optimal parameter
  selection for the alternating direction method of multipliers ({ADMM}):
  quadratic problems,'' \emph{IEEE Trans. Autom. Control}, vol.~60, no.~3, pp.
  644--658, Mar. 2015.

\end{thebibliography}

\end{document}